\documentclass[11pt]{article}
\usepackage[utf8]{inputenc}
\usepackage[T1]{fontenc}
\usepackage{bm}
\usepackage{amsmath,amsfonts,amssymb,multirow}
\usepackage{amsthm}
\usepackage{multicol}
\usepackage{appendix}
\usepackage{cite}

\usepackage[margin=1in]{geometry}
\usepackage{epsfig}
\usepackage{graphics}
\usepackage{float}
\usepackage[dvipsnames,usenames]{color}
\usepackage[colorlinks=true,urlcolor=Blue,citecolor=Green,linkcolor=BrickRed]{hyperref}
\usepackage[usenames,dvipsnames]{xcolor}
\usepackage{algorithm}
\usepackage[noend]{algorithmic}

\usepackage{paralist}
\usepackage{todonotes}

\usepackage{enumitem}

\newtheorem{claimm}{Claim}
\newtheorem{lemma}{Lemma}

\newtheorem{observation}{Observation}
\newtheorem{property}{Property}

\makeatletter

\renewcommand{\theenumi}{\arabic{enumi}}

\renewcommand{\p@enumii}{\theenumi.}
\makeatother

\newcommand{\DDG}{\mathit{DDG}}

\makeatletter

\begin{document}

\title{Minimum Cut of Directed Planar Graphs in $O(n\log\log n)$ Time \thanks{The research was supported in part by Israel Science Foundation grant 794/13.}}

\author{
\!\!\!\!\!\! Shay Mozes \\
\!\!\!\!\!\! IDC Herzliya \\
\!\!\!\!\!\! \href{mailto:smozes@idc.ac.il}{smozes@idc.ac.il}   
\and \!\!\!\!\!\!
Cyril Nikolaev \\ \!\!\!\!\!\! University of Haifa\\ \!\!\!\!\!\!
\href{mailto:cyril.nikolaev@gmail.com}{cyril.nikolaev@gmail.com} \\
\and \!\!\!\!\!\!
Yahav Nussbaum \\ \!\!\!\!\!\! University of Haifa\\ \!\!\!\!\!\!
\href{mailto:yahavnu@gmail.com}{yahavnu@gmail.com}
\and \!\!\!\!\!\! Oren Weimann \\
\!\!\!\!\!\! University of Haifa\\  \!\!\!\!\!\!\href{mailto:oren@cs.haifa.ac.il}{oren@cs.haifa.ac.il}
}

\date{}
\maketitle

\begin{abstract}
	We give an $O(n \log \log n)$ time algorithm for computing the
        minimum cut (or equivalently, the shortest cycle) of a
        weighted directed planar graph. This improves the previous
        fastest $O(n\log^3 n)$ solution. 
           Interestingly, while in undirected planar graphs both min cut and min $st$-cut have $O(n \log \log n)$ solutions, 
 in directed planar graphs our result makes 
              min cut faster than min $st$-cut, which currently requires $O(n \log n)$.
        
\end{abstract}

\newpage
\clearpage
\setcounter{page}{1}

\section{Introduction}

A \emph{cut} is a partition of the vertex set of a graph into two
non-empty sets $X$ and $Y$. The \emph{capacity} of a cut is the
total capacity of the edges from $X$ to $Y$. The \emph{minimum cut}
  problem asks to find a cut with minimum capacity. 
The \emph{minimum $st$-cut}
  problem asks, in addition, that vertex $s$ belongs to $X$ and
vertex $t$ to $Y$. In undirected planar graphs, both 
problems can be solved in $O(n
\log\log n)$-time~\cite{INSWN11,LS11}, where $n$ is the number of vertices of the graph. In directed planar graphs,
however, the fastest algorithms currently known run in $O(n \log^3n)$
for min cut~\cite{WN09}, and in $O(n\log n)$ for min
$st$-cut~\cite{BK09}. In this work we show how to find a  min cut in a
directed planar graph in $O(n
\log\log n)$ time. Therefore, we can currently solve min cut  faster than min
$st$-cut in directed planar graphs.

There is a well
known duality between cuts in a planar graph and cycles in the
\emph{dual planar graph}. A minimum cut in a planar
graph is  a \emph{shortest cycle} in the dual planar
graph. It follows that  any algorithm for finding the minimum cut in a
planar graph can also find the shortest cycle in a planar graph, and vice versa.

\vspace{-10pt}
\paragraph{\bf Undirected planar graphs.} For an {\em undirected}
planar graph $G$, Chalermsook, Fakcharoenphol and Nanongkai~\cite{CFN04} gave a
simple algorithm that finds the 
minimum cut
by recursively separating the dual graph $G^*$ with {\em shortest path
separators}. At
each recursive step, the CFN algorithm applies a min $st$-cut
algorithm\footnote{Actually, CFN~\cite{CFN04}
 use a max $st$-flow algorithm. They
  cite~\cite{Weihe},  but
  that algorithm is flawed, as was pointed out by Borradaile and
  Klein~\cite{BK09}, who also gave a correct algorithm.} in $O(n\log n)$ time. This gives an $O(n \log^2 n)$-time
algorithm for undirected 
min cut.
Improvements to this running time are based on using faster min
$st$-cut algorithms in the CFN algorithm. One such
algorithm is that of Reif~\cite{R83}, which is a divide and conquer
algorithm over a shortest $s$-to-$t$ path. 
We refer to
this algorithm as the {\em shortest}-path based algorithm.
Italiano et al.~\cite{INSWN11} showed how to use a technique by
Fakcharoenphol and Rao~\cite{FR06} to implement the shortest-path
based algorithm in $O(n \log \log n)$ time. Plugging this into the CFN
algorithm yields an $O(n \log n \log \log n)$ time algorithm
for undirected 
min cut~\cite{INSWN11}. 

A second min $st$-cut algorithm in undirected planar graphs is that of
Kaplan and Nussbaum~\cite{KaplanN11}. This algorithm is based 
on a divide and conquer algorithm on an $s$-to-$t$ path that is not necessarily a shortest path, but is small in terms of the number of its vertices. We refer to this algorithm as the {\em small}-path based
algorithm.
In the paper mentioned above, Italiano et al.~\cite{INSWN11} 
used the fact that the small-path based algorithm runs in sublinear
time when the small path has a sublinear number of vertices, in order to design a min
$st$-cut {\em oracle} with sublinear query time and $O(n \log\log n)$
preprocessing time.
\L\k{a}cki and Sankowski~\cite{LS11}  showed how to efficiently
represent and maintain the shortest path separators and the information required by the
small-path based min $st$-cut oracle of Italiano et al. along the execution of the CFN algorithm.
This allowed them to implement each of the $O(\log n)$
recursive levels of the CFN algorithm in sublinear time. The overall running time is $O(n \log\log n)$, which is the current state of
the art for min cut in undirected planar graphs. Note that, in
undirected planar graphs, both min cut and min $st$-cut
currently take $O(n \log\log n)$ time.

\vspace{-10pt}
\paragraph{\bf Directed planar graphs.} The min cut in directed planar graphs, as noted in~\cite{WY10}, can be  found in $O(n^{3/2})$ time with a simple use of planar separators. Wulff-Nilsen~\cite{WN09} used the afore-mentioned technique of
Fakcharoenphol and Rao, to bring the running time down to
$O(n\log^3n)$, which is the fastest algorithm to this problem prior to
the current work. For min $st$-cut in
directed planar graphs, the fastest known algorithm is the $O(n\log n)$-time max $st$-flow
algorithm of Borradaile and Klein~\cite{BK09}.
Note that in the directed case there is a gap between the $O(n
\log n)$-time algorithm for min $st$-cut~\cite{BK09} problem, 
and  the  min cut problem which, until the present work, required
$O(n \log^3 n)$ time.

\paragraph{Our results and techniques.} 
In this paper, we present an $O(n \log \log n)$ time algorithm for finding
the minimum cut (and hence also the shortest cycle)
in a weighted directed planar graph. We believe this is a significant advance on
a fundamental optimization problem.

First, we make a simple observation, that was somehow overlooked, showing that the structural lemma underlying 
 the $O(n \log^2 n)$-time CFN algorithm~\cite{CFN04} (for min cut in undirected planar
graphs) can actually be proven for the directed case as well. 
It is then easy to modify the CFN algorithm to work for directed planar
graphs in the same complexity; 
In undirected graphs, a
minimum cut separating $s$ and $t$ is a minimum $st$-cut. In directed graphs it may be a $ts$-cut.
We therefore 
compute both a min $st$-cut and a min $ts$-cut at each step of the recursion of the CFN algorithm. The running time of the
algorithm remains $O(n \log^2 n)$.

Recall that improving upon the CFN algorithm in
 the undirected case required faster min $st$-cut algorithms.
However, both the shortest-path based algorithm, 
and the small-path based algorithm, 
rely heavily on the graph being undirected. 
Consequently, it seems that getting faster algorithms for directed min
$st$-cut is very difficult, and that, therefore, any progress on the minimum cut problem in
directed planar graphs is unlikely.
Surprisingly, we show this is not the case. 
We make another simple observation which bypasses
this difficulty. We show that, while the shortest-path based 
min $st$-cut algorithm does not work in the directed setting, it does
work in the directed setting when the min $st$-cut happens to be the global minimum cut!
Though simple, this surprising observation is a main conceptual contribution of our work. 
This observation alone immediately  implies that Italiano
et al's $O(n \log\log n)$ implementation of the shortest-path based  min $st$-cut algorithm~\cite{INSWN11} can be used in
the CFN algorithm to find the min cut in directed planar
graphs in $O(n \log n \log\log n)$ time.

Getting the running time down to $O(n \log\log n)$ turns out to be
much more difficult. The small-path based algorithm, on which
\L\k{a}cki and Sankowski's algorithm is based, heavily relies on
the graph being undirected, and we do not know how to use it in the
directed setting, even for finding the global min cut.
Instead, we develop an implementation of the CFN algorithm  that uses 
the shortest-path based algorithm, rather than the
small-path one. In this implementation, each recursive step takes sublinear {\em
  amortized} time rather than worst case time as in \L\k{a}cki and
Sankowski's. We believe this yields a somewhat simpler
 algorithm, even for undirected min cut, since the small-path based min $st$-cut oracle is quite
complicated. 

The most technically involved part of our contribution is in overcoming the difficulties that arise when combining the efficient implementation
of the shortest-path based algorithm a la Italiano et al. with the implicit
representation of \L\k{a}cki and Sankowski. 
The result is the first directed
variant of Reif's algorithm and the first to handle non-simple directed cycles.
An interesting component in our solution is the use of auxiliary 
non-planar (but bounded genus) graphs. This allows us to guarantee that certain subpaths 
that are represented implicitly posses structural properties that are required for the correctness of our algorithm.
It is often the case that algorithms for planar graphs are used in
algorithms for bounded genus graphs. Here an
algorithm for bounded genus graphs is used for solving a problem on
planar graphs. We find this use very interesting.
An overview of the difficulties and their resolution can be found in Section~\ref{sec:holes}.

Beyond making significant progress on a fundamental
optimization problem using an interesting and technically challenging 
solution, our result puts the landscape of planar minimum cut problems in
an interesting situation.  Whereas undirected minimum cut, undirected
minimum $st$-cut and directed minimum cut can all be solved in $O(n
\log\log n)$, we only know how to compute directed minimum
$st$-cut in $O(n \log n)$ time. This may hint that the algorithms for 
min $st$-cut and max $st$-flow in directed planar graphs
can also be improved.

\vspace{-10pt}
\paragraph{Bounded genus graphs.} For bounded genus graphs, 
some of the algorithms above~\cite{D10,E99,WY10,WN09} work with a minor modification. In particular, it is possible to show that, on a weighted directed graph with genus $g$, the algorithm of Djidjev~\cite{D10} finds the shortest cycle in $O(g^{1/2}n^{3/2})$ time, and the algorithm of Wulff-Nilsen~\cite{WN09} in $O(gn \log^2 n + n \log^3 n)$ time.
We show how to use ideas from our planar algorithm to find a 
shortest cycle in a graph of genus $g$ in
$O(g^2 n \log n)$ time with high probability or $O(gn\log^2n)$ time in the worst case. 

\section{Preliminaries} \label{sec:pre}
In this section we provide necessary background and definitions. Most
of the material covered is not new. However, this section does contain
a number of novel insights and observations that are original
contributions of this work. These are clearly indicated where appropriate.

\vspace{-10pt}
\paragraph{Basic concepts.}
We assume basic familiarity with planar graphs, such as familiarity
with the definition
of the planar dual and the duality of cuts and cycles.
Let $G$ be a simple directed planar graph with $n$ vertices and non-negative
arc weights. 
A directed path $P$ is a sequence of arcs  $P = v_0 v_1, v_1 v_2, \dots,
v_{k-1} v_k$. It is a directed cycle if, in addition,  $v_0 = v_k$.
An \emph{undirected} path (cycle) is a sequence of edges such that reorienting
some of the edges yields a directed path (cycle).
Unless otherwise stated, all paths and cycles are directed. 
We write $u <_P v$ to denote that vertex $u$ appears before vertex $v$
in the path $P$.
We denote by $P[u,v]$ the subpath of $P$ starting at vertex $u$ and
ending at vertex $v$, and by $P(u,v)$, the subpath $P[u,v]$ without
its first and last edges. 
Also, $P[\cdot,a]$ ($P[a,\cdot]$) denotes the prefix (suffix) of $P$
ending (starting) at $a$.
We denote by $rev(uv)$ the arc $vu$, and by $rev(P)$ the path whose
arcs are the reverse of the arcs of $P$ in reverse order. 
We denote the number of arcs on path $P$ by $|P|$. The length of $P$
is the sum of lengths of $P$'s arcs.  
 
\begin{figure}[h]
	\centering
	\includegraphics[scale=.4]{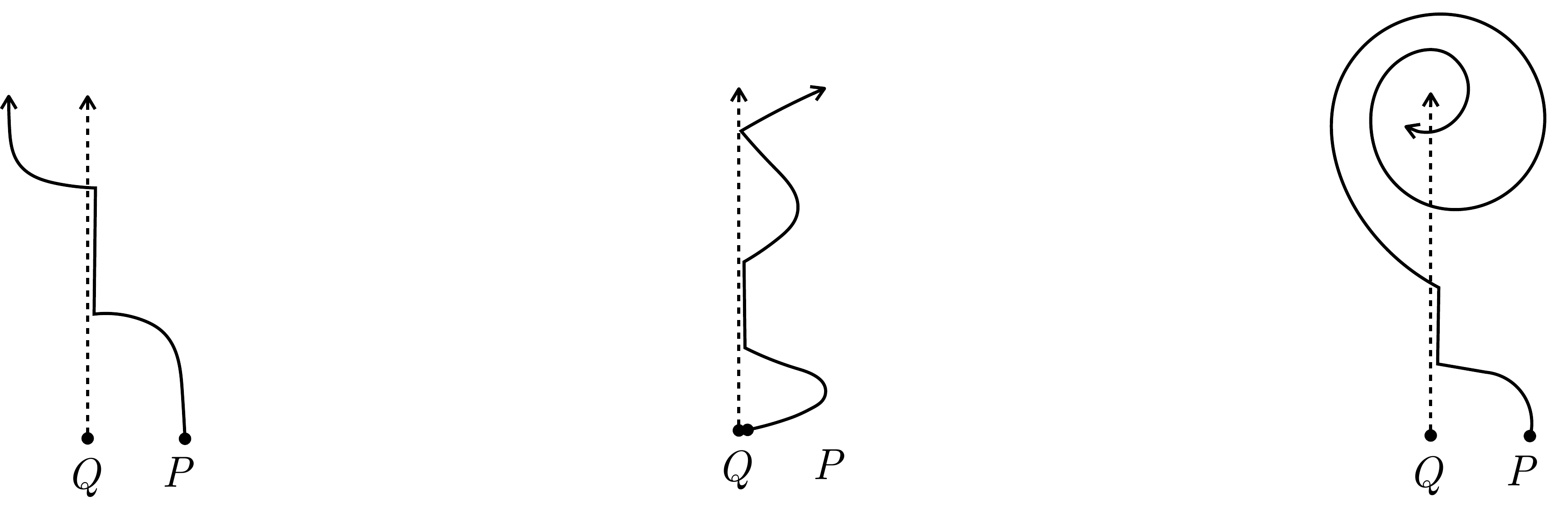}
	\caption{On the left, the path $P$ ({\em solid}) crosses the path $Q$ ({\em dashed}). On the middle, $P$ does not cross $Q$. On the right, $P$ crosses $Q$ three times (all from right to left) so their crossing parity is odd.}
	\label{fig:cross}
\end{figure}

We say that a path $P$ \emph{crosses} another path $Q$ if there is a
path $R$ that is a common subpath of $P$ and $Q$ such that (i) the
first (last) vertex of $R$ is not the first (last) vertex of $P$ or
$Q$, and (ii) the edge of $P$ that precedes the subpath $R$ enters $Q$
from one side and the edge of $P$ that follows $R$ leaves $Q$ from the
other side. See Figure~\ref{fig:cross}. The absolute value of the
number of times $P$ crosses $Q$ from right to left minus the number of
times $P$ crosses $Q$ from left to right is called the {\em crossing
  number} of $P$ and $Q$. Its parity is called the the {\em crossing
  parity}.   
The crossing number (parity) of a primal path $P$ and a dual path $Q$ is
defined as the (parity of the) number of arcs of $P$ whose duals belong
to $Q$ minus the number of reverses of arcs of $P$ whose duals belong
to $Q$.

We say that a (possibly non-simple) cycle $C$ {\em encloses} a face $f$ if a path starting at a
virtual vertex embedded in the infinite face and ending at a virtual
vertex embedded in $f$ crosses $C$ an odd number of times. A vertex or
an edge $x$ incident to a face $f$ are enclosed by $C$ if $f$ is enclosed by $C$. If $x$ is enclosed by $C$ but $x \not \in C$ then
$x$ is said to be {\em strictly enclosed} by $C$.
The subgraph enclosed by a cycle $C$ is called the {\em interior} of $C$ and
the subgraph not enclosed by $C$ is called the {\em exterior} of $C$ ($C$
itself belongs to both the interior and the exterior).

\vspace{-10pt}
\paragraph{Unique shortest paths.}
We assume throughout the paper that shortest paths in the graph are
unique. This assumption simplifies the presentations, but is also used
in proving the correctness of our algorithm.
In general graphs, this assumption can be achieved with high
probability by applying the Isolation
Lemma~\cite{Isolation,Isolation2}. Indeed, prior algorithms for
ebedded graphs that require this assumption usually use the isolation
lemma, which results in a randomized algorithm with high probability of
success. However, recently, Erickson and Fox~\cite{Private} have
shown a simple way to enforce this assumption deterministically in
graphs embedded on a genus $g$ surface with $O(g)$ overhead (i.e., with no overhead for planar graphs).
  
\vspace{-10pt}
\paragraph{Multiple-source shortest paths (MSSP).}

Klein~\cite{K05} described an algorithm that, given a directed planar
graph $G$ with arc lengths, a face $f_\infty$ of $G$, and a shortest
path tree $T$ rooted at some vertex of $f_\infty$ computes, in
$O(n\log n)$ time, a data structure representing  all shortest path trees rooted
at each vertex of $f_\infty$. The data structure can be queried in
$O(\log n)$ time for the distance between any vertex $u \in
f_\infty$ and any other vertex $v \in V(G)$. The data structure can
also be queried for the arcs of the shortest $u$-to-$v$ path (instead of just the
distance), in $O(\log \log \Delta)$ amortized time per reported arc. Here
$\Delta$ is the maximum degree of a vertex in $G$. We refer to this
algorithm and data structure as MSSP (multiple-source shortest
paths). 
Cabello, Chambers and Erickson~\cite{CCE13} described an MSSP algorithm for
genus-$g$ graphs. The algorithm assumes unique shortest paths and runs in $O(g n\log n)$ time with high
probability (using the isolation lemma), or in deterministic $O(g^2
n\log n)$ time (using
the new technique of Erickson and Fox~\cite{Private}).

\vspace{-10pt}
\paragraph{\boldmath$r$--divisions, Dense distance graphs, and FR-Dijkstra.}
An $r$-{\em division}~\cite{F87} of $G$, for some  $r < n$, is a decomposition of $G$
 into $O(n/r)$ pieces,  
where each piece has at most $r$ vertices and $O(\sqrt{r})$ \emph{boundary} vertices (vertices shared with other pieces).
There is an $O(n)$ time algorithm that computes an $r$-division 
of a planar graph with the additional property
that the boundary vertices in each piece lie on a constant number of 
faces of the piece (called {\em holes})~\cite{FR06,KMS13}.
The \emph{dense distance graph} (DDG) of a piece $R$ is the complete graph over the boundary vertices of $R$. The length of edge 
$uv$ in the DDG of $R$ equals to the $u$-to-$v$ distance inside $R$. Note that the DDG of $R$ is non-planar. The DDG of an $r$-division is the union of DDGs of all pieces of the $r$-division. 
Thus, the total
number of vertices in the DDG is sublinear $O(\frac{n}{r} \cdot \sqrt r) =
O(\frac{n}{\sqrt r})$, and the total
number of edges is linear $O(\frac{n}{r} \cdot r) = O(n)$.
The DDG can be computed in $O(n\log r)$ time  
using the MSSP algorithm~\cite{K05}.
Fakcharoenphol and Rao~\cite{FR06} described an implementation of
Dijkstra's algorithm, nicknamed {\em FR-Dijkstra} on the DDG of an
$r$--division. 
Computing shortest paths in the DDG using FR-Dijkstra takes
$O(\frac{n}{\sqrt r} \log^2 (\frac{n}{\sqrt r}))$ time which is
proportional (up to polylog factors) to the number of vertices of the
DDG, and sublinear in $n$, the number of vertices of $G$.

\vspace{-10pt}
\paragraph{A directed version of the CFN algorithm.}\label{sec:dividingstep} 
The algorithm of Chalermsook et al.~\cite{CFN04} computes a minimum
cut in an undirected planar graph.  We describe their algorithm for the directed
case.\footnote{The observation that the CFN algorithm can be made to work in
the directed case is novel.} For this we need the following lemma, which implies that we may assume that the shortest cycle in the graph
crosses any shortest path at most once.\footnote{A similar lemma
appeared without proof in~\cite{CFN04}, but that paper did not
consider directed graphs.}

\begin{lemma} \label{lem:commonsubpath}
 	Let $P$ be a shortest $u$-to-$v$ path for a pair of vertices
        $u$, $v$. There is a globally shortest cycle $\mathcal
        C$ such that
        either $\mathcal C$ and $P$ are completely disjoint or they share  a  single subpath.
\end{lemma}

Let $o$ be an arbitrary vertex in $G$. A \emph{shortest path
  separator}~\cite{LT79} is an undirected cycle $S$ consisting of an
edge $uv$, a shortest (directed) $o$-to-$u$ path $P$, and a shortest
(directed) $o$-to-$v$
path $P'$, such that both the interior and exterior of the cycle
consist of at most $2/3$ of the total number of the faces of $G$. 
Such a separator can be found in $O(n)$ time~\cite{LT79,CFN04}.

\begin{figure}[h]
	\centering
	\includegraphics[scale=.6]{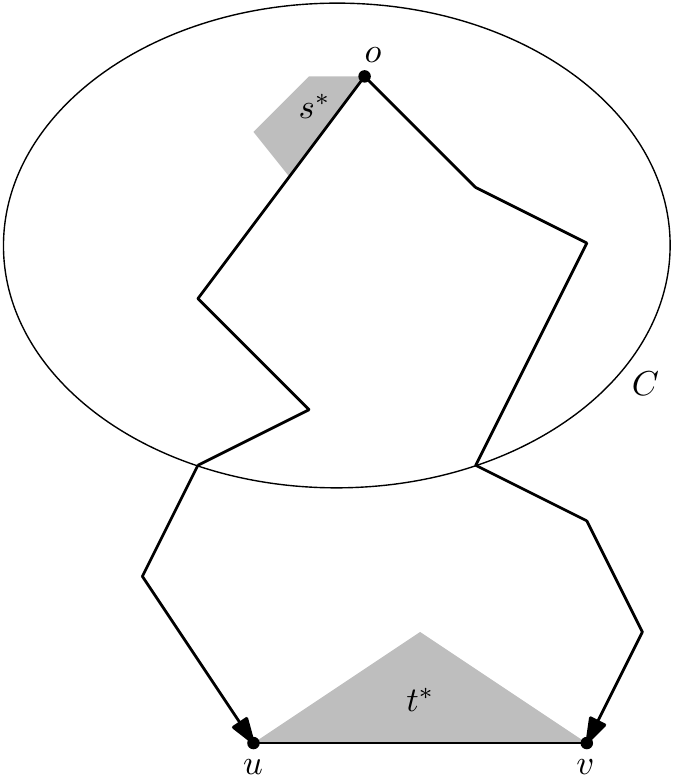}
	\caption{If the shortest cycle $C$ crosses the separator $S$, then it crosses the shortest $o-u$ path once and the shortest $o-v$ path once. In this case, the vertex $o$ (together with the face $s^\ast$) and the edge $uv$ (together with the face $t^\ast$) are in two different sides of $C$.}
	\label{fig:conquer}
\end{figure}

Given a shortest path separator $S$, the shortest cycle in $G$ is either in the interior of  $S$, in the exterior of $S$, or
crosses $S$. The former two options are handled recursively. We
describe how to find the shortest cycle $\mathcal C$ that crosses $S$. 
Since $\mathcal C$ crosses $S$, it does so at least twice. By
Lemma~\ref{lem:commonsubpath}, we may assume that $\mathcal C$ crosses
$P$ exactly once, and so 
the vertex $o$ and the edge $uv$ are in two different sides of
$\mathcal C$. 
Let $s^\ast$ be the face adjacent to the first edge of $P$ external to  $S$,
and let $t^\ast$ be the face adjacent to $uv$ internal to $S$. The cycle
$\mathcal C$ is the shortest cycle that separates $s^\ast$ and $t^\ast$. See Figure~\ref{fig:conquer}. 
In the \emph{dual planar graph}, $\mathcal C$ is either a minimum
$st$-cut or a minimum $ts$-cut, where $s$ and $t$ are the vertices
dual to the faces $s^\ast$ and $t^\ast$, respectively. Therefore,
$\mathcal C$ can be found by two executions of a min $st$-cut algorithm, which takes $O(n \log n)$ time~\cite{BK09}. 

Overall, the recursive decomposition of the graph using shortest path
separators has $O(\log n)$ levels of recursion. Before each recursive level we can remove
every vertex of degree two, and merge its two adjacent edges into a
single edge (combining the lengths of the two). This guarantees that
the total size of all subgraphs in the same  level of the recursion is $O(n)$,
and so all executions of the min $st$-cut algorithm in this level take
total $O(n \log n)$.  The overall running time is thus $O(n \log^2 n)$.

\vspace{-10pt}
\paragraph{Reif's algorithm.}
\label{sec:conquerstep}
Reif's algorithm~\cite{R83} (referred to in the introduction as the
shortest-path based algorithm) is used to find a minimum
$st$-cut in an undirected planar graph. 
We describe it as an algorithm
for a directed graph $G$.\footnote{We note that the literature is infested with
  inaccuracies on the use of Reif's algorithm in the directed case. Janiga and Koubek~\cite{JK92}
  attempted to generalize Reif's algorithm to compute a min $st$-cut
  in directed planar
  graphs. They find the shortest cycle that separates $s^\ast$ and
  $t^\ast$ and crosses some $s$-to-$t$ path at a particular
  vertex. This cycle is used  to divide the problem into two separate
  subproblems. However, this algorithm is
  flawed~\cite{KaplanN11}. The cycle found by Janiga and Koubek  is
  not necessarily 
  simple, nor do they make sure it corresponds to an $st$-cut rather
  than to a $ts$-cut. Erickson and Nayyeri~\cite{EN11} remarked that the algorithm
  of Janiga and Koubek appears to find the smaller between the minimum
  $st$-cut and minimum $ts$-cut. However, this claim is also
  false. The cycle dual to the min $st$-cut may cross the 
  cycle used by Janiga and Koubek to divide the problem. 
In  this case the cycle corresponding to the min $st$-cut will never be
  found because it does not belong to any of the two subproblems.}
Given a shortest $s^*$-to-$t^*$ path $P$ in $G$, Reif's algorithm
finds the shortest cycle $C$ that crosses $P$ exactly
once.\footnote{This view of Reif's algorithm does not require deep
  insights but is novel nonetheless. It does require a slightly careful
  proof of Lemma~\ref{lem:reif}, which is trivial in the undirected
  case.} 
In undirected graphs $C$ corresponds to a min $st$-cut in the dual graph, but in directed
graphs it does not. The crucial observation, however, is that this is
exactly the property required by the CFN algorithm for finding the global min cut. 

\begin{figure}[h!]
	\centering
	\includegraphics[scale=0.6]{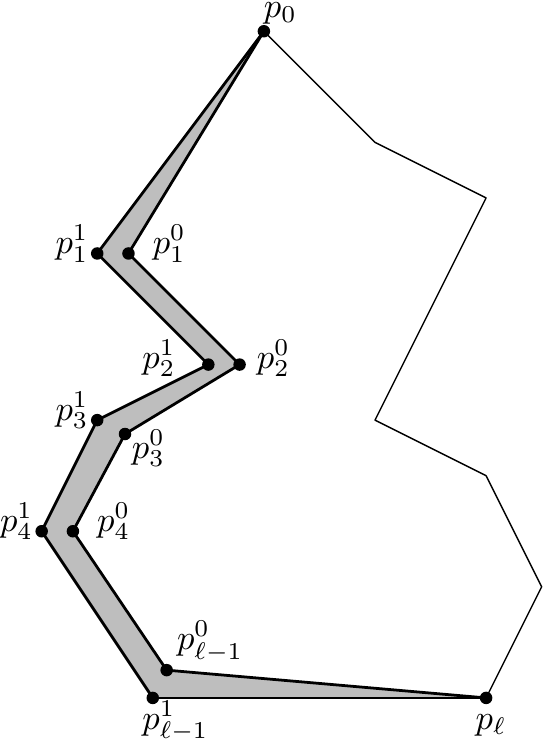}
	\caption{An incision along $P$. The newly created face is \emph{shaded}.}
	\label{fig:incision}
\end{figure}

We assume that the cycle $C$ crosses the path $P$ from right-to-left
(the other case is symmetric, and the algorithm tries both). Reif's algorithm makes an \emph{incision} along $P$ and replaces every vertex $p_i$ of $P$ with two vertices $p_i^0$ and $p_i^1$. 
Every edge $p_ip_{i+1}$ of $P$ is replaced with two edges
$p_i^0p_{i+1}^0$ and $p_i^1p_{i+1}^1$. Every edge $p_iv$ is replaced
with an edge $p_i^0v$ ($p_i^1v$) if it emanates left (right) from
$P$. Similarly, every edge $vp_i$ is replaced with an edge $vp_i^0$ ($vp_i^1$) if it enters $P$ from its left (right) side. See Figure~\ref{fig:incision}.

Let $P_i$ be a shortest $p_i^0$-to-$p_i^1$ path. In the original graph
(i.e., before  the incision) $P_i$ is a shortest simple cycle
$C_i$ which crosses $P$ exactly once, at $p_i$. Finding the desired
cycle $C$ therefore amounts to finding the shortest among all $P_i$s.
 Reif's algorithm does this in $O(n \log n)$ time using
 divide-and-conquer based on the following lemma.

\begin{lemma} \label{lem:reif}
  Let $P_i$ be a shortest $p_i^0$-to-$p_i^1$ path.
	For $j \neq i$, there is a shortest $p_j^0$-to-$p_j^1$ path $P_j$ that does not cross $P_i$.
\end{lemma}

Reif's divide-and-conquer algorithm proceeds as follows. It first
finds a shortest $p_i^0$-to-$p_i^1$ path $P_i$ for $i =  |P|/2$. This
takes $O(n)$ time using~\cite{HKRS97}. 
The path $P_i$ divides the graph into two
subgraphs. By Lemma~\ref{lem:reif}, each subgraph can be handled
separately. The algorithm therefore recurses on both subgraphs.
To get a total running time of $O(n \log n)$, in each recursive level
we remove vertices of degree two and merge their two adjacent edges as
explained in the CFN algorithm above.

\vspace{-10pt}
\paragraph{Italiano et al's implementation of Reif's algorithm.}\label{sec:Italiano}
Italiano et al.~\cite{INSWN11} developed a faster $O(n \log \log n)$
implementation of Reif's algorithm to find a minimum $st$-cut in
undirected planar graphs. 
As above, we observe that, when applied to a directed
planar graph, the algorithm computes the shortest  simple cycle
crossing $P$ exactly once. Plugging this into each of the $\log n$ levels of the
CFN algorithm yields  an $O(n \log n \log \log n)$ algorithm for the directed global
min cut problem.

The algorithm of Italiano et al. computes an $r$-division with $r =\log^6n$.
As in Reif's algorithm, an incision is made in $G$ along $P$. Note that the
incision may cut pieces.   
Every such piece $R$, is  replaced with a set of pieces, one for each
connected component of $R$ following the incision. For every vertex
$p_i$ of $P$ that was a boundary vertex prior to the incision,  both
$p_i^0$ and $p_i^1$ are defined to be boundary vertices after the incision. 
The DDG of all resulting pieces can be computed 
in $O(n \log r) = O(n \log \log n)$ time using the MSSP
algorithm~\cite{K05}, and we can run FR-Dijkstra on
this DDG in {\em sublinear} $O((n / \sqrt{r}) \log^2 n) = O(n / \log n)$ time.

The first stage of the algorithm (called {\em coarse Reif}) finds the
shortest cycles $C_i$ that cross $P$  once {\em at a boundary vertex}.
The running time of this step is dominated by the $O(n \log\log n)$ time required to
compute the DDG. It implements Reif's algorithm by only considering
boundary vertices, and uses FR-Dijkstra to quickly compute the shortest paths $P_i$.
The next step, called {\em refined Reif}, computes the shortest cycles
that cross $P$ at non-boundary vertices. It implements Reif's
algorithm within the subgraphs enclosed by the cycles found in the
coarse Reif step. The running time of this step is also $O(n\log\log n)$.

Italiano et al.\ used the main ideas from their fast implementation of Reif's
algorithm to design a  min $st$-cut oracle for undirected planar
graphs that, after $O(n\log\log n)$
preprocessing can answer min $st$-cut queries, and
support edge insertions and deletions, in 
$O(n/\log n)$ time per query or operation. This oracle is based on a
min $st$-cut due to Kaplan and Nussbaum~\cite{KaplanN11}, rather than
on Reif's. The oracle was then used by \L\k{a}cki and Sankowski~\cite{LS11} to solve undirected
  global min cut as we explain next.

\vspace{-10pt}
\paragraph{The algorithm of \L\k{a}cki and Sankowski for undirected
  global min cut.}
The currently fastest algorithm for undirected global min cut is that
of \L\k{a}cki and Sankowski~\cite{LS11}. Its running time is  $O(n
\log\log n)$. 
Their algorithm emulates the CFN
algorithm on the DDG.
The bottleneck in the $O(n \log n \log \log n)$ global min cut algorithm of Italiano et al.~\cite{INSWN11}
is the recomputation, in $O(n \log\log n)$ time, of the DDG at each of the $O(\log n)$ levels of the
CFN recursion.
\L\k{a}cki and Sankowski~\cite{LS11} showed how  to build the DDG just once (in $O(n \log\log n)$ time)
and maintain it (in sublinear time) throughout all the recursive calls
of the CFN algorithm.
They further show how to find a shortest path separator in $O(n/\log n)$ time,
and maintain the information required by the min cut oracle of Italiano et al.\ to compute min
$st$-cuts in $O(n/\log n)$ time. Thus, the running time of the whole algorithm is
actually dominated by the $O(n \log\log n)$-time preprocessing step of building the DDG.

\L\k{a}cki and Sankowski described how
to efficiently keep track of the partition of the 
graph into subgraphs when cutting along a cycle $C_i$ that is only
represented in the DDG (this is called implicitly cutting the
graph). 
The vertices of the DDG (i.e., the boundary vertices of $G$)
are partitioned into the interior and exterior of $C_i$ according to
the embedding of $C_i$ in $G$. The time required is proportional to
the number of boundary vertices, not to the size of $G$. We use this
technique in our algorithm without change. 
A brief description of the so called recursion graph and division edges used in
their technique appears in Section~\ref{sec:division-edges} for completeness. 

\section{Our Algorithm} \label{sec:ouralgo}
Our  observations from the previous section allow us to design a
version of the CFN algorithm that is based on an efficient directed 
variant of Reif's algorithm. 
Our algorithm begins by computing an $r$-division of the graph
$G$, and a corresponding DDG for $r=\log^8 n$. This takes $O(n \log r) = O(n \log\log
n)$ time.
Then, as in~\cite{LS11}, a shortest path tree of $G$, rooted at some boundary vertex, is computed and maintained as a shortest
path tree in the DDG. The dividing step identifies 
a balanced shortest path separator composed of two shortest paths $P$ and $P'$ plus a single
edge $e$. Let $B=\{b_1,\ldots b_p\}$ be the boundary vertices along the
shortest path $P$. Since $P$ ends at a vertex
  of $e$, which is not necessarily a boundary vertex, the suffix
  $P[b_p,\cdot]$ is fully contained in the piece of the $r$-division
  containing $e$. 
The algorithm represents $P$ by the sequence of boundary vertices $B$ plus all the vertices in the suffix  $P[b_p,\cdot]$.
Note that $P$ may have $O(n)$ vertices but its representation uses only 
  $O(n/\sqrt r)$ boundary vertices, and the $O(r)$
 vertices of $P[b_p,\cdot]$. 
The algorithm cuts the graph along the separator, as done
in~\cite{LS11},  and recurses on the interior
and exterior subproblems.\footnote{Note that, since the graph is
  directed, edges of the DDG in a subproblem may represent paths that
  cross the separator an
even number of times. This does not affect the correctness of the
algorithm because we are interested in the globally minimum cycle. 
On the one hand, such paths are at least as short as
the shortest path restricted to the subproblem. On the other hand, such
paths correspond to valid paths in the original graph.}
Problems with fewer than $r$ boundary
vertices are not handled recursively, but by any existing directed
global min cut algorithm (see analysis). In addition, we invoke a global
min cut algorithm on every piece $R$ individually.

In the conquering step, we wish to find
the shortest cycle that crosses $P$ exactly
once. This is where our algorithm significantly differs
from~\cite{LS11}. Instead of using the min $st$-cut oracle of~\cite{INSWN11}, we present 
a directed variant of Reif's algorithm (which we refer to as the {\em inner} recursion).
In what follows we assume, without loss of generality, that the shortest cycle $\mathcal
C$ we are looking for crosses $P$ from right to left. The other case
is symmetric, and the algorithm implements both.

\vspace{-10pt}
\paragraph{Performing an incision along  \boldmath$P$.}\label{section:incision}
We now describe the procedure for performing an incision along $P$
(the preliminary step of Reif's algorithm).
Consider a piece $R$. If any subpath of $P$ connects two different
holes of $R$ or if $t$ is a vertex of $R$ then we perform the incision of $R$ explicitly. 
Otherwise, the incision is performed implicitly. In an explicit
incision, a piece $R$ is explicitly cut into subpieces, and a
DDG is computed from scratch for each of the resulting subpieces by rebuilding their MSSP data structure~\cite{K05}. In
an implicit incision, the edges of the DDG of $R$ are partitioned among
the DDGs of the subpieces of $R$, without actually cutting $R$ and
recomputing the DDG. This is done as follows. The
subpaths of $P$ going through $R$ break $R$ into connected
components. See Figure~\ref{fig:h-cyclic-cover}(left).
Each of these connected components is considered from now
on as an individual {\em subpiece} of $R$. 
The division of the boundary vertices of $R$ (on all holes of $R$)
into the resulting subpieces is inferred as in~\cite{LS11}, using the
skeleton graph and the division edges technique.
For each subpiece $Q$ we would like the length of each
DDG edge $uv$ to correspond  to the shortest $u$-to-$v$
path in $Q$ (rather than in $R$). However, this would require recomputing the DDG of $Q$
which we cannot afford. Instead, we use the original DDG edge $uv$ in
$R$. This edge corresponds to a shortest $u$-to-$v$ path $\rho$ that is allowed to
venture in $R$ outside $Q$. It turns out that this is problematic only
when the region $R$ contains holes. 
For ease of presentation we ignore this point for now and proceed with
describing the algorithm. We will later discuss the
 difficulties manifested by holes and their 
resolution.

\vspace{-10pt}
\paragraph{Applying Reif's algorithm to \boldmath$P$.}
Having made the incision along $P$ in the DDG, we now wish to perform
Reif's divide-and-conquer on the path $P$. However, since an edge of
the graph might appear on the path $P$ in many different levels of the
CFN recursion, we cannot afford to handle all edges of $P$ at every
recursive level. We next show that it suffices to handle each edge $e$ at most once, at the earliest level of the CFN recursion at which $e\in P$. 

We classify the edges of $P$ into 
 two types, {\em active} and {\em
  inactive}. An edge $p_ip_{i+1}$ is inactive if it was already part
of the separator in some earlier level of the CFN recursion.
Observe
that (1) the active edges form a suffix of $P$, (2) we only need to
find the shortest $p_i^0$-to-$p_i^1$ path $P_i$ if $p_ip_{i+1}$ is
active (if $p_ip_{i+1}$ is inactive then 
every cycle that goes through $p_i$
  must also go through $p_{i+1}$), and (3) we can discover the active
  suffix by revealing the edges of $P$ one by one (each in $O(\log r)$
  time using the MSSP data structure
until we reach an inactive edge. 

The first step of our Reif variant therefore discovers the active suffix of $P$ in time $O(x\log r)$ where $x$ is the number of active edges in $P$. 
Next, we wish to  
find the shortest $p_i^0$-to-$p_i^1$ path $P_i$ where $p_i$ is the middle vertex of the active suffix of $P$.  
Let $R$ denote the piece containing $p_i$. We (temporarily) add $p_i^0$ and $p_i^1$ as  boundary
vertices and add appropriate DDG edges as follows. If there exists a subpath of $P$ whose endpoints lie on different holes of $R$,
it means that we have already explicitly built the new DDG of $R$'s
subpieces (after the incision) by computing new MSSP data structures
(see above). 
In this case $p_i^0$ and $p_i^1$ both belong to the same subpiece $Q$,
we add to the DDG an edge from $p_i^0$ to every boundary vertex of $Q$, and from every boundary vertex of $Q$ to $p_i^1$. The lengths of these edges are obtained by querying the new MSSP data structure of $Q$. 
Otherwise, the endpoints of all subpaths of $P$ in $R$ lie on the same hole so $p_i^0$ and $p_i^1$ belong to different subpieces $Q_0$ and $Q_1$ of $R$. We add to the DDG
edges from $p_i^0$ to all vertices of $Q_0$ and from all vertices of $Q_1$
to $p_i^1$.
These distances are computed by querying 
the existing MSSP data structures of $R$. 

After connecting $p_i$ to the boundary vertices of $R$, we find the DDG representation of the path $P_i$ by running FR-Dijkstra from $p_i^0$ to $p_i^1$ on the DDG. Notice that while $P_i$ is a simple path in the DDG, it might correspond to a non-simple path in the underlying graph. This is because the implicit DDG incision means DDG edges may correspond to subpaths in the graph before the incision. We later show how to ensure that this does not violate the correctness of the algorithm. 

Reif's algorithm proceeds by cutting the graph along the cycle $C_i$
that corresponds to $P_i$ and recursing on the interior and
exterior. We implement this by cutting the DDG implicitly  along  $P_i$ using the division edge
technique, obtaining two DDGs denoted $\DDG_s$ (containing $s$) and
$\DDG_t$ (containing $t$). We assign the prefix 
$P[s,p_i)$ to $\DDG_s$  and the suffix $P(p_i,t]$ to $\DDG_t$, and then
recurse on both subgraphs.

We note that, for {\em undirected} graphs, the algorithm above is complete and correct. We believe that, for the undirected case, it is simpler than the algorithm of \L\k{a}cki and Sankowski~\cite{LS11} because it does not rely on the rather complicated min $st$-cut oracle of Italiano et al.~\cite{INSWN11}.  

\subsection{A flaw in the algorithm and its resolution}\label{sec:holes}
Since our algorithm implements the CFN algorithm, in order to prove
the correctness of our algorithm it suffices to prove that, at each
level of the CFN recursion, if the
global min cycle $\mathcal C$ crosses the shortest path separator then
our algorithm will find $\mathcal C$. Since at each level of the CFN recursion our
algorithm implements Reif's algorithm, it suffices to show that any
$p_i^0$-to-$p_i^1$ path $P_i$ found by our algorithm is either
$\mathcal C$, or  $\mathcal C$ is represented in one of the resulting
 $DDG_s$ or $DDG_t$ obtained by implicitly cutting the DDG along
$P_i$. However, in directed graphs this
might actually be false! We first explain at a high level how the problem arises, and then explain how to resolve it.

\begin{figure}[h!]
	\centering 
	\includegraphics[scale=.51]{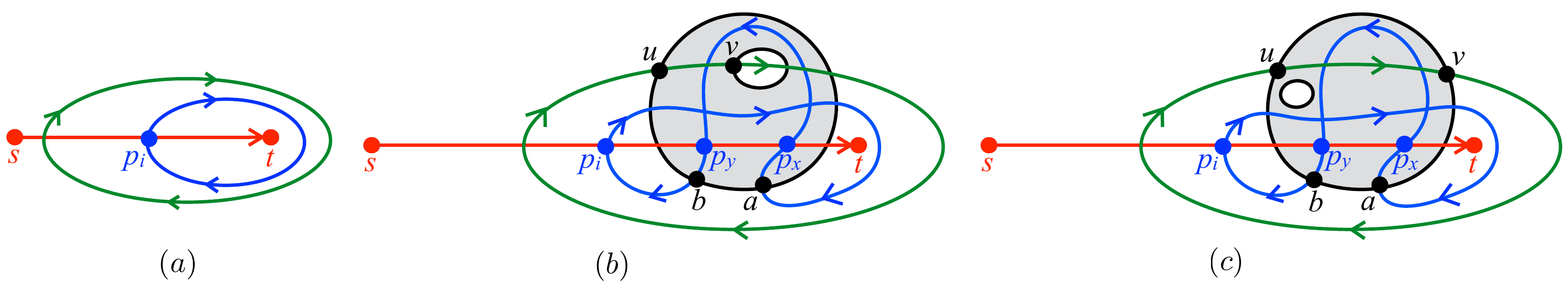}
	\caption{\small{ 
In all diagrams, the shortest $s$-to-$t$ path $P$ is
          shown in red, the global shortest cycle $\mathcal
          C$ in green, and a the
          shortest cycle $C_i$ that crosses $P$ at the middle vertex
          $p_i$ in blue. ($a$) When $C_i$ is simple it is not crossed by
          $\mathcal C$, so breaking the problem along $C_i$ is
          valid. ($b$) Even though $P_i$ is a simple path in the DDG, the
          cycle $C_i$ it corresponds to in the underlying graph (blue) is not simple. A piece $R$ of the
          $r$-division is shown (shaded). The underlying path of the
          DDG edge $ab$ crosses $P$ even after an implicit incision along
          $P$. The $p_x$-to-$p_y$ subpath of $P_i$ is called a
          finger. In this case, the globally shortest cycle $\mathcal C$
          might cross $C_i$ at the finger. Dividing the problem
          along $C_i$ is problematic because $\mathcal C$ has boundary
          vertices in both the exterior (e.g., $u$) and the interior
          (e.g., $v$) of $P_i$.  ($c$) When no holes are
          ``sandwiched'' between the finger and $P$, all boundary
          vertices of the globally shortest cycle $\mathcal C$ are in
          the exterior of $C_i$ even though $\mathcal C$ crosses
          $C_i$. Therefore, $\mathcal C$ is still represented in $DDG_s$
          after breaking the problem implicitly along $C_i$.}
	\label{fig:intro}
}
\end{figure}

Recall that the DDG of a subpiece $Q$ used at some point in the execution of
our algorithm is not obtained by explicitly computing shortest paths
between the boundary nodes in that subgraph, but by using edges from
the DDG of the original piece $R$. This implies that the shortest paths
that correspond to DDG edges of $Q$ may actually venture outside $Q$. 
In particular, while the $p_i^0$-to-$p_i^1$ path $P_i$ found by our
algorithm is a simple cycle in the DDG that crosses $P$ exactly once
(at $p_i$), it may actually correspond to a non-simple cycle $C_i$ that
crosses $P$ more than once. See Figure~\ref{fig:intro}(b,c).
It turns out that in such cases $\mathcal C$ may actually cross
$C_i$. This is problematic because the algorithm implicitly cuts the
DDG along $P_i$ and recurses on $DDG_s$ and $DDG_t$. If $\mathcal C$
crosses $P_i$ then it seems that $\mathcal C$ will be represented  in
neither $DDG_s$ nor $DDG_t$.

To overcome this problem we characterize the structure of the subpaths
of $P_i$ that cross $P$. We call such subpaths {\em fingers}. We show that
each finger is restricted to a single piece $R$ of the
$r$--division. We further show that the problem mentioned above does
not occur in fingers that enclose no holes of $R$ (see
Figure~\ref{fig:intro}(c)). The reason is that, in the absence of
holes, all boundary vertices on the global min cut do lie on the same
side of $C_i$. Therefore, if a finger encloses no holes, even though $\mathcal C$ might
cross $C_i$, it is still represented in either $DDG_s$ or $DDG_t$.
Since the number of holes in each piece
is constant, we can precompute a constant number of versions of the
DDG of each piece $R$. In each version, the paths corresponding to DDG
edges interact with the holes of $R$ in a prescribed way. We can then
carefully choose which version of the DDG  of $R$ to use when computing $P_i$ so as
to ensure that each finger of $C_i$ is locally homologous to $P$ in roughly the
following sense: the subgraph sandwiched between the finger and $P$ contains
no holes of $R$.

We now explain the changes in the algorithm in detail.
Recall the description of the implicit incision along $P$. 
Let $R$ be a piece of the
$r$--division. The path $P$ breaks $R$ into subpieces. 
For each subpiece $Q$ we would like the length of each
DDG edge $uv$ to correspond  to the shortest $u$-to-$v$
path in $Q$ (rather than in $R$). However, this would require recomputing the DDG of $Q$
which we cannot afford. Instead, we use the original DDG edge $uv$ in
$R$. This edge corresponds to a shortest $u$-to-$v$ path $\rho$ that is allowed to
venture in $R$ outside $Q$.
The path $\rho$ can be decomposed so that
the maximal subpaths of $\rho$ in $R \setminus Q$ start and end on
$P$. We call these subpaths {\em simple fingers} of $\rho$. The {\em base} of a
simple finger is the subpath of $P$ between the endpoints of the finger. For
the correctness of our algorithm we
require that:
\begin{property}\label{prop:simple-finger}
For every simple finger $S$, the cycle formed by $S$ and its
base encloses no holes of $R$.	
\end{property}
 To achieve Property~\ref{prop:simple-finger}, instead of precomputing a
single DDG for each piece $R$ of the $r$-division, we compute many DDGs
(exponential in the number of holes in $R$, which is $O(1)$). When
information about a DDG edge $uv$
of $Q$ is required (e.g., by FR-Dijkstra or when implicitly cutting
the graph open), it is reported using the
precomputed version of the DDG of $R$ that corresponds to the subset
of holes that belong to $Q$. We next explain this preprocessing step. 

\begin{figure}[h]
	\centering
	\includegraphics[scale=.8]{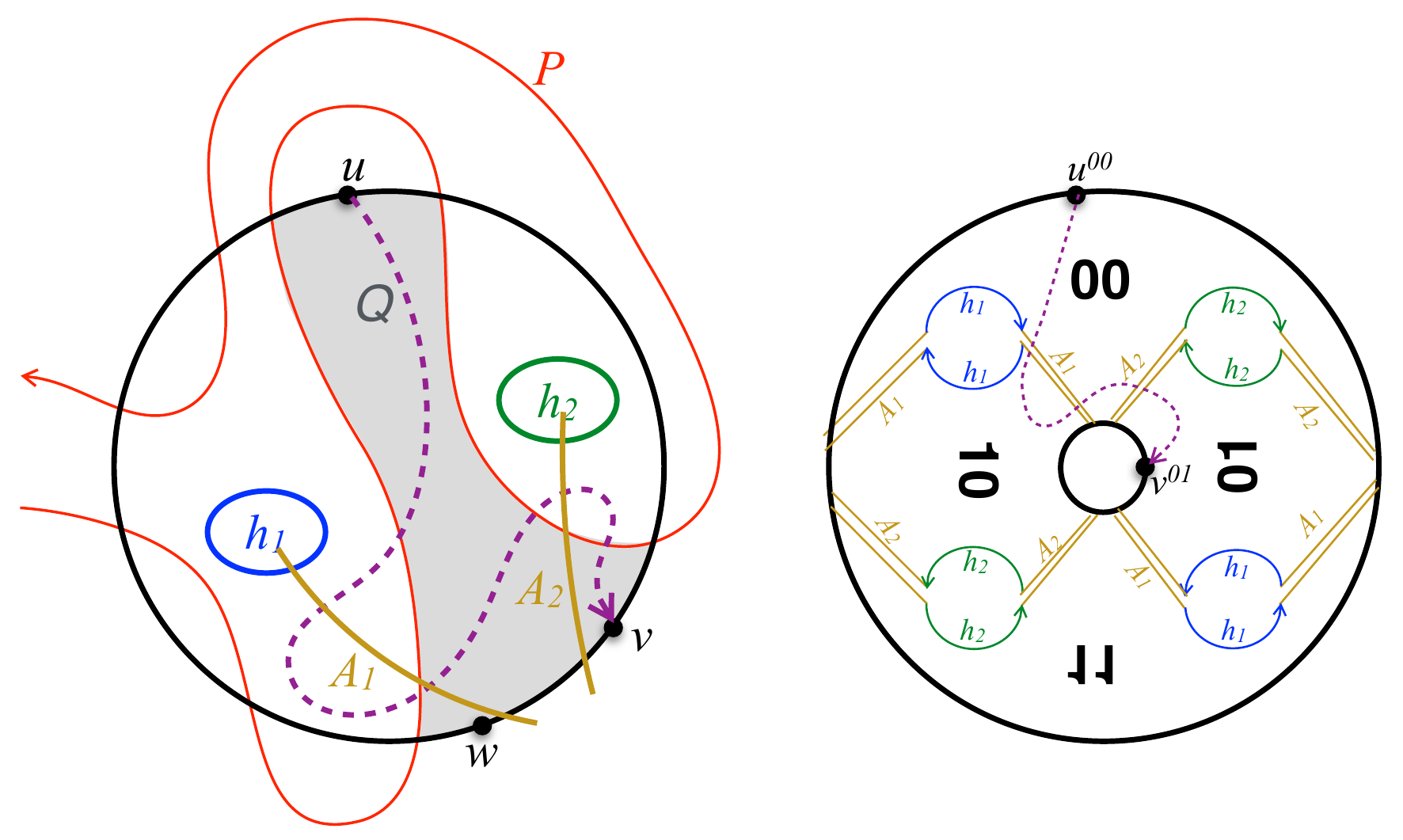}
	\caption{On the left, a piece $R$ (in black) with two holes
          $h_\ell$ ($\ell=1,2$, green and blue), each with a dual path $A_\ell$
          from the external hole to $h_\ell$. The path $P$ (in red) separates $R$ into subpieces, one of which (shaded) is $Q$. On the right, the $\mathbb{Z}_2$-homology cover of $R$ with $H=\{h_1,h_2\}$. A valid (i.e., one whose fingers do not enclose any holes of $H$) shortest $u$-to-$v$ path $\rho$ in $R$ must have an even (odd) crossing parity with $A_1$ ($A_2$) and therefore corresponds to a $u^{00}$-to-$v^{01}$ path in the $\mathbb{Z}_2$-homology cover. One such valid $u$-to-$v$ path $\rho$ is illustrated as the dashed (purple) line. 
Similarly, a valid $u$-to-$w$ path must have an odd (even) crossing parity with $A_1$ ($A_2$) and therefore corresponds to a $u^{00}$-to-$w^{10}$ path in the $\mathbb{Z}_2$-homology cover. 
}\label{fig:h-cyclic-cover}
\end{figure}

\vspace{-10pt}
\paragraph{\bf  The $\pmb{\mathbb{Z}_2}$-homology cover.}
We use a special case of the $\mathbb{Z}_2$-homology cover developed
by Erickson and Nayyeri~\cite{EN11} for bounded genus graphs. To the
best of our knowledge this is the first time that homology covers are used for planar graphs. 
Our description is less general
than the one in~\cite{EN11}, and differs in some
of the details to make the presentation shorter and suitable for our
application.  We perform the following preprocessing for each of the $O(1)$
possible subsets $H$ of holes of $R$. 
For each hole $h_\ell \in H$ we choose an arbitrary path $A_\ell$ in the dual
of $R$ connecting the external hole of $R$ with $h_\ell$. 
 We construct a graph, called the $\mathbb{Z}_2$-homology
cover of $R$ by making, for each $\ell=1, \dots, |H|$, an incision
along $A_\ell$. See section~\ref{sec:conquerstep}, and
Figure~\ref{fig:incision} for a detailed definition of an
incision. Note, that here the incision is performed in the dual of
$R$.  In the primal, this can be thought of as splitting each (primal) edge of
$A_\ell$ into two complementary half-edges that are not connected to each other. See
Figure~\ref{fig:h-cyclic-cover}. Let $R'$ denote the resulting (primal)
graph. The $\mathbb{Z}_2$-homology cover is constructed by glueing
together $2^{|H|}$ copies of $R'$. Each copy is labeled with a distinct $|H|$-bit string. For
labels $b$ and $b'$ differing in a single bit $j$, the corresponding
copies of $R'$ are glued along the complementary half edges of $A_j$. See Figure~\ref{fig:h-cyclic-cover}.
The resulting graph is not planar,
but can be embedded on a surface with constant (albeit exponential in $|H|$)
genus.\footnote{The dual of this graph is essentially a hypercube (which has genus $2^{|H|}$). That is, after deleting from each copy of $R'$ all edges that do not belong to the infinite face, the interior of each copy becomes a single face and the dual of this $\mathbb{Z}_2$-homology cover is a hypercube. Adding back the deleted (planar) portions does not increase the genus. }
We can therefore use the MSSP data structure for bounded genus
graphs~\cite{CCE13} on the $\mathbb{Z}_2$-homology cover of
$R$ in $O(r \log r)$ time. 
Let $B_Q$ be the set of boundary vertices of $Q$ that do not
belong to holes in the subset $H$.
The MSSP data structure 
can report in $O(\log r)$ time the distance between
any vertex $u^{0\dots 0}$ of $B_Q$ (i.e., the boundary vertex $u$ in the copy with the all-zero
label) and any vertex $v^{b}$ of $B_Q$ (i.e., the boundary vertex $v$ in the copy
with label $b$). A shortest $u^{0\dots 0}$-to-$v^b$ path in the $\mathbb{Z}_2$-homology cover corresponds to a shortest $u$-to-$v$ path in $R$ under the
restriction that for every $A_\ell$ it's crossing parity is even (odd)
iff the $\ell$th bit of $b$ is zero (one).  See
Figure~\ref{fig:h-cyclic-cover}.
When information about a DDG edge $uv$ of $Q$ is required during the
execution of the algorithm, it is fetched by querying the MSSP data
structure for the $\mathbb{Z}_2$-homology cover created for the
appropriate subset $H$ of the holes of $R$. For the query  
we need to figure out the label $b$ of $v$
to be used when querying the MSSP data structure. This choice is described in
Section~\ref{sec:more-cover}.

\subsection{Correctness} 

The correctness of our algorithm follows from the following lemma, which states that in the way we cut the DDG we do not lose the globally shortest cycle $\mathcal C$.

\begin{lemma}\label{lem:correct}
	If the globally minimum cycle $\mathcal C$ is the 
 $p_k^0$-to-$p_k^1$ path in the DDG for some $p_k \in P$, 
then either $p_k=p_i$ and  $\mathcal C = C_i$, or $p_k \in P[s,p_i)$ and  $\mathcal C$ is the
$p_k^0$-to-$p_k^1$ path in $\DDG_s$, or $p_k \in P(p_i,t]$ and   $\mathcal C$ is the
$p_k^0$-to-$p_k^1$ path in $\DDG_t$.
\end{lemma}

\noindent	In the rest of this section we lay out the structural
properties that facilitate the proof of Lemma~\ref{lem:correct}. The
proof itself is a rather complicated case analysis and is deffered to
Section~\ref{sec:horribleproof}. In a nutshell, our analysis 
shows that, in every possible case, either $\mathcal C$
crossing $C_i$ leads to a contradiction or the crossing is such that
all the boundary vertices of $\mathcal C$ (and hence all
the DDG edges of  $\mathcal C$) belong to either $\DDG_s$
or $\DDG_t$.

Observe that in the DDG both $\mathcal C$ and $C_i$ cross $P$ exactly
once (from right to left). In the underlying graph however, $C_i$ may
cross $P$ some odd number of times. The cycle $C_i$ can be partitioned
into internally disjoint subpaths that do not cross $P$ at all. See
Fig.~\ref{fig:Correctness} for an illustration.
	 Starting from $p_i$, $C_i$ is first composed of zero or more
         alternating $p_x$-to-$p_y$ subpaths where $y < x < i$ (otherwise, if $x < y$ then by the unique shortest paths assumption $C_i[p_x,p_y]$ should be equal to
         $P[p_x,p_y]$). These  subpaths either begin by emanating left of
         $P$ and end by entering left of $P$ or they begin by
         emanating right of $P$ and end by entering right of $P$. We
         call the former subpaths a {\em finger of $C_i$ above $P$}
         and the latter a {\em finger of $C_i$ below $P$}. 
	 After such zero or more fingers, there is exactly one $p_\ell$-to-$p_j$  subpath that begins by emanating left of $P$ at $p_\ell$ and ends by entering right of $P$ at $p_j$. We call this $p_\ell$-to-$p_j$ subpath the {\em separation finger}.  
	 Observe that by definition $\ell\le i$. 
	 If $j<i$, then it must be (by the unique shortest paths assumption) that $C_i[p_j,p_i] = P[p_j,p_i]$. Otherwise, if $j\ge i$, then  $C_i[p_j,p_i]$ consists of zero or more alternating $p_x$-to-$p_y$ subpaths where $i< y < x < j$. These  subpaths can be fingers of $C_i$ above $P$ or below $P$. 
	 
	  Note that, because the DDG was implicitly cut along $P$, for
          every DDG edge $uv$ of $C_i$ that belongs to a piece $R$
          such that $P$ separates $R$ into multiple parts, both $u$
          and $v$ belong to the same part $Q$. If the path
          corresponding to the DDG edge $uv$ crosses $P$, it must do
          so an even number of times, and create at least one finger
          that  belongs to $R \setminus Q$. recall that any such
          finger is called a simple finger. Hence:
	           
         \begin{observation}\label{observation:below}
         Every $p_x$-to-$p_y$ finger $S$ of $C_i[p_i,p_\ell]$ below $P$ is a simple finger. Thus $S$ is contained in a single piece of the $r$-division and (by Property~\ref{prop:simple-finger}) $S$ encloses no holes.
                    \end{observation}
         
         \begin{observation}\label{observation:above}
         Every $p_x$-to-$p_y$ finger $S$ of $C_i[p_j,p_i]$ above $P$ is a simple finger. Thus $S$ is contained in a single piece of the $r$-division and (by Property~\ref{prop:simple-finger}) $S$ encloses no holes.

         \end{observation}
    
Note that by our incision procedure we have that fingers of $C_i$ that are confined to a single piece $R$ do not enclose any holes of $R$. Also note that two fingers of $C_i$ can cross each other
(thus making $C_i$ non-simple). However, by the following claim, this can only happen if one finger is the separation finger and the other is
a finger of $C_i[p_j,p_i]$ above $P$ (Figure~\ref{fig:Correctness}) or a finger of $C_i[p_i,p_\ell]$ below $P$ (Figure~\ref{fig:Correctness3}), or one finger is of $C_i[p_i,p_\ell]$ and the other is of $C_i[p_j,p_i]$ (Figure~\ref{fig:Correctness2}).  

\begin{claimm}\label{claim:crossing}
$C_i$ can cross itself only if the crossing is (I) between the separation finger and 
a finger of $C_i[p_j,p_i]$ above $P$, or (II) between the separation finger and 
a finger of $C_i[p_i,p_\ell]$ below $P$, or (III) between a finger of $C_i[p_i,p_\ell]$ above (below) $P$ and a finger of $C_i[p_j,p_i]$ above (below) $P$.
\end{claimm}

\noindent The remainder of the proof of Lemma~\ref{lem:correct} which consists of numerous cases appears in Section~\ref{sec:horribleproof}.

\subsection{Analysis}

All pre-computations are dominated by the $O(n\log r) = O(n\log\log n)$ time for computing the DDG in all pieces of the $r$--division.
 
We first bound the time for performing incisions.
The time to perform the implicit incisions of $P$ is proportional to
the number of boundary vertices in the graph, which is $O(n/\sqrt r)$ over all
subgraphs in a single level of the CFN recursion tree. Hence, over the
entire run of the algorithm, the cost of all implicit incisions is
$O((n /\sqrt r)\cdot \log n) = O(n)$.
The time to perform an explicit incision of a single piece $R$ is
dominated by the 
$O(r\log r)$ time of MSSP. Every time such an incision is made because
a DDG edge $e$ of $P$ connects two different holes of $R$, the number of holes in
$R$ decreases (the two holes connected by $e$ become a single
hole). Since the number of holes in each piece is constant, such explicit incisions occur
a constant number of times per piece over the entire execution of the
algorithm. Hence, the total time spent on all such explicit incisions over
all pieces during the entire course of the algorithm is $O((n/r)\cdot r\log
r)  = O(n\log\log n)$. 

In each subproblem of the CFN recursion, the algorithm performs an explicit incision in the piece $R$ to which $t$ (the last
vertex of $P$) belongs. Such incisions do not decrease the number of
holes, so we cannot charge for them globally as above. In subproblems
containing $\Omega(r)$ boundary vertices, the $O(r \log r)$ time of
the implicit incision is dominated by the $O(r \log^2 n)$ time for FR-Dijkstra
computation. Subproblems with $O(r)$ boundary vertices are called {\em
  small subproblems} and are handled
by running any existing algorithm for directed min cut (i.e.,
bootstrapping). 
Denote the running time
of this directed min cut algorithm by $O(n f(n))$. Then, as
shown in~\cite{LS11}, handling  
all small subproblems takes $O(n f(r))$ time.  

We now bound the time spent on FR-Dijkstra computations.
Consider a non-small subproblem at some level of the CFN recursion with $b=\Omega(r)$ boundary vertices and $x$ active edges. The algorithm finds
$C_i$ using FR-Dijkstra in $O(\sqrt r \log r + b \log^2 n)$
time (the first term is the cost of connecting the middle vertex $p_i$
to the
DDG, and the second term is the cost of FR-Dijkstra). The algorithm then recurses on
the interior and exterior of $C_i$ each containing at most $x/2$
active edges. Contracting degree-2 edges in the DDG of every subproblem  guarantees that, at each level of the inner
recursion, every boundary vertex appears in at most two 
subproblems with more than two boundary vertices~\cite{INSWN11,LS11}. Call a subproblem with at most two boundary vertices a \emph{tiny} subproblem. Thus, the total time for all non-tiny subproblems along all levels
of the inner recursion is $O(b \log^2n \log x)$, which is $O(b \log^3n)$ since $x=O(n)$. Since the sum over the boundary vertices of all non-small subproblems at a single level of the CFN recursion is $O(n/\sqrt
r)$,  and since the depth of the CFN recursion is $O(\log
n)$, the total time required for all FR-Dijkstra computations on
non-tiny subproblems throughout the algorithm is $O(n\log^4 n/\sqrt r )= O(n)$.

We now bound the cost of tiny subproblems. This cost is dominated by the $O(\log r)$ time required to connect $p_i$ to the (at most two) boundary vertices. The Dijkstra computation then takes constant time. Since each tiny subproblem is associated with some active edge (an edge whose endpoint is $p_i$), and since at this time this edge becomes inactive, the total number of tiny problems along the entire execution of the algorithm is $O(n)$, and the time to handle them all is $O(n\log r) = O(n \log\log n)$.  

Subproblems with no boundary vertices are handled by  bootstrapping. I.e., by calling any directed min cut algorithm in each piece of the $r$--division separately. This takes total $O((n/r)\cdot  rf(r) = O(n f(r))$ time.

Summing the different terms above we get that the total running time of the
algorithm is $O(n \log\log n) + O(n f(r)) +O(n) + O(n \log\log n)  + O(n f(r)))$. Using the $f(n) =
\log^2n$ (i.e., using~\cite{CFN04} for bootstrapping) results in a
total running time of $O(n \log\log n + n \log^2(\log n)) = O(n
\log^2\log n)$ for directed global minimum cut. Using the resulting
algorithm in the bootstrapping (i.e., $f(n) = \log^2\log n$) yields the
claimed $O(n \log\log n)$ running time for directed global minimum cut.

\section{Missing Proofs and Additional Details}
\subsection{Proof of Lemma~\ref{lem:commonsubpath}}
	Let $\mathcal C$ be a shortest cycle in $G$ that shares two
        distinct vertices $c_1$ and $c_2$ with the path $P$, labeled
        so that  $c_1 <_P c_2$. If the subpath of $\mathcal C[c_1,c_2]$ is different than $P[c_1,c_2]$, then we replace  $\mathcal C[c_1,c_2]$ with $P[c_1,c_2]$. Since $P$ is a shortest path, the cycle $\mathcal C$ remains a shortest cycle. We repeat this process until the vertices and the edges of $\mathcal C$ that are also in $P$ form a subpath of $\mathcal C$, as required. 
\subsection{Proof of Lemma~\ref{lem:reif}}
\begin{figure}[h!]
\centering
		\includegraphics[scale=0.65]{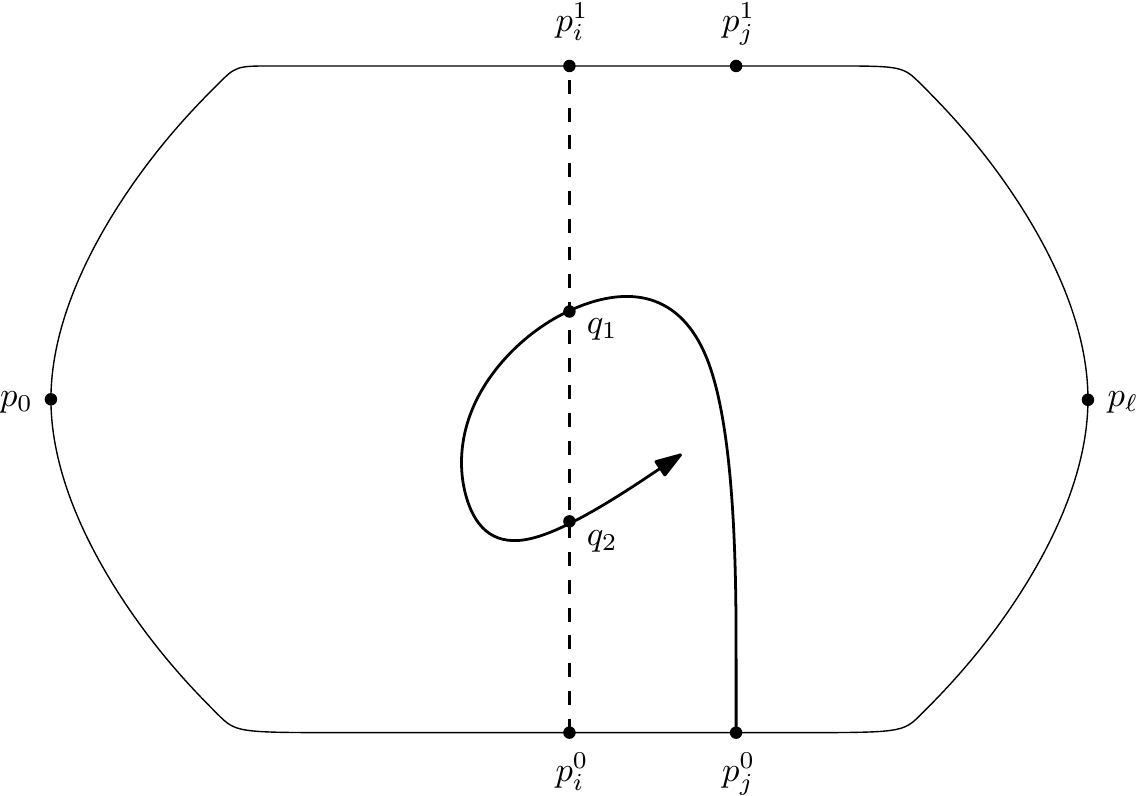}
 		\caption{The path $P_i$ (\emph{dashed}) separates the graph into two subgraphs. Both $p_j^0$ and $p_j^1$ are in the same side of $P_i$. If the path $P_j$ (\emph{solid}) crosses $P_i$ such that the first vertex of $P_j$ that is common with $P_i$ follows in $P_i$ the last vertex of $P_j$ that is also a vertex of $P_i$, then $P_j$ must cross itself. Note that, for clarity, we re-embedded the graph so that the new face created by the incision is the infinite face.}
 		\label{fig:reif}
\end{figure}

   The path $P_i$ separates the graph into two subgraphs. The two
         vertices $p_j^0$ and $p_j^1$ are in the same side of
         $P_i$. Let $P_j$ be a simple shortest $p_j^0$-to-$p_j^1$
         path. If $P_j$ crosses $P_i$ then $P_j$ must touch $P_i$
         after the first such crossing
since the two endpoints of $P_j$ are in the same side of
         $P_i$. Let $q_1$ be the first vertex of $P_j$ that also
         belongs to $P_i$, and let $q_2$ be the last vertex of $P_j$
         that also belongs to  $P_i$. It must be that  $q_1 <_{P_i} q_2$
         since otherwise $P_j$ must cross itself (see
         Figure~\ref{fig:reif}). The subpath of $P_i[q_1,q_2]$ is a
        shortest $q_1$-to-$q_2$ path. Replacing $P_j[q_1,q_2]$ with
         $P_i[q_1,q_2]$ results in a shortest $p_j^0$-to-$p_j^1$ path $P_j$ that does not cross $P_i$, as required.

\subsection{Proof of Claim~\ref{claim:crossing}}
\begin{figure}[h]
	\centering 
	\includegraphics[scale=.6]{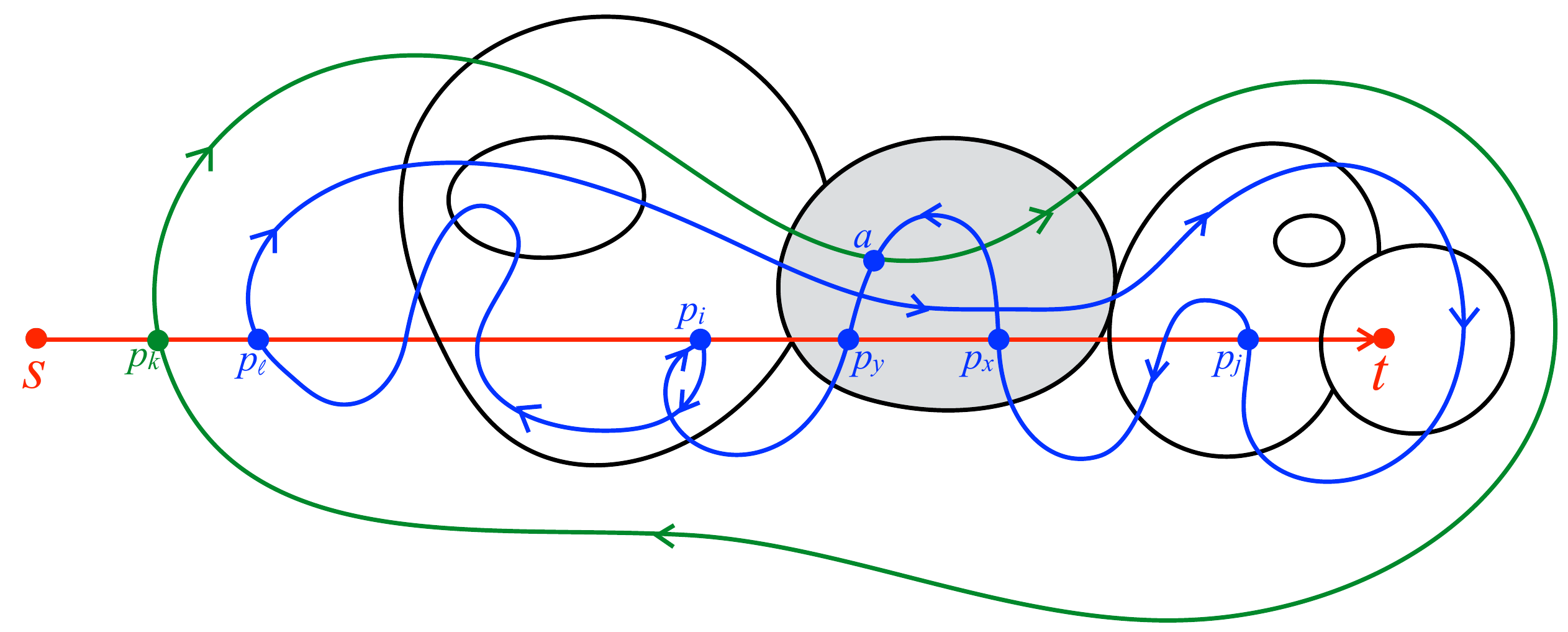}
	\caption{The shortest $s$-to-$t$ path $P$ (in red), some pieces of the $r$-division (in black), the shortest cycle $C_i$ (in blue), and the globally shortest cycle $\mathcal C$ (in green). 	Apart from the $p_\ell$-to-$p_j$ separation finger, $C_i$ has four fingers below $P$ and three above.  
	Observe that $\mathcal C$ does not cross $C_i$ except inside the shaded piece where $\mathcal C$ enters and exits
	a $p_x$-to-$p_y$ finger of $C_i$ above $P$. Such crossings are allowed.}
	\label{fig:Correctness}
\end{figure}

We prove that all other crossings are impossible since they imply that we can remove a subcycle $C'$ from $C_i$ (thus making $C_i$ shorter) while $C_i$ still passes through $p_i$ and its crossing parity with $P$ remains the same:
\begin{itemize}
\item A finger cannot cross itself. This is because apart from its endpoints a finger does not include any vertices of $P$ and so if it crosses itself at vertex $a$ it means that there is a cycle $C'$ containing $a$ but not containing any vertices of $P$.  

\item Two fingers cannot cross if they are both above $P$ or both below $P$ and are both in either $C_i[p_j,p_i]$ or $C_i[p_i,p_\ell]$. If they cross at vertex $a$ then there is a cycle $C'$ that (1) contains $a$, (2) does not contain $p_i$, and (3) crosses $P$ an even number of times (since $C'$ does not contain $C_i[p_\ell,p_j]$). 

\item The separation finger cannot cross a finger of $C_i[p_i,p_\ell]$ above $P$ or a finger of $C_i[p_j,p_i]$ below $P$ for the same reason as the previous case.  

\item A $p_x$-to-$p_y$ finger $f_1$ below $P$ and a $p_w$-to-$p_z$ finger $f_2$ above $P$ cannot cross. Assume for contradiction that $f_1$ first crosses $f_2$ at vertex $a$. Consider the cycle $C''$ composed of (1) the subpath of $f_1$ between $p_x$ and $a$, (2) the subpath of $f_2$ between $a$ and $p_z$, and (3) the subpath of $P$ between $p_x$ and $p_z$. Notice that at vertex $a$, $C_i[p_x,\cdot]$ enters $C''$ and must exit $C''$ before reaching $p_y$. It cannot exit at (1) because we proved that a finger $f_1$ cannot cross itself, and it cannot exit at 
(3) because a finger does not cross $P$, so it must exit $C''$ at some vertex $b$ of $f_2$ that belongs to $C_i[a,p_z]$. However, this means that both $f_1$ and $f_2$ contain $a$-to-$b$ subpaths in contradiction to unique shortest paths.
\end{itemize}

\subsection{Proof of Lemma~\ref{lem:correct}}
\label{sec:horribleproof}
To prove the lemma, we next show that at least one of $\DDG_s$ or
$\DDG_t$ contains all the boundary vertices of $\mathcal C$. 
Suppose for the sake of contradiction that there are two boundary
vertices on $\mathcal C$ that do not belong to the same side of $C_i$. 
Let $p_k$ be the vertex where $\mathcal C$ crosses $P$.
Let
$b_1, b_2$ be the first pair of consecutive boundary vertices on
$\mathcal C$ after $p_k$ that belong to different sides of $C_i$.

Assume w.l.o.g. that $p_k$ and $b_1$ are on the same side of $C_i$.
Let $a$ be the first vertex of $\mathcal C$ after $b_1$ where $\mathcal C$ crosses $C_i$. 
First we consider the case where $p_k$ is strictly external to $C_i$. The finger $S$ containing $a$ is one of three types:
\begin{enumerate}[leftmargin=*]

\item {\em $S$ is the separation finger.} In this case, since $p_k$ is
  external to $C_i$,  there can be three options: (1) $j\le k < \ell$,
  (2) $k <j$ and $k<\ell$, (3) $k  \ge \ell$.
	
	\begin{enumerate}

	\item  If $j\le k < \ell$, then $C_i[p_j,p_i]= P[p_j,p_i]$ and $p_k$ is a vertex of $C_i$, so it is not strictly external to $C_i$.   

	\item \label{case1.2} If  $k <j$ and $k\leq\ell$, then let $C'$ be the cycle $\mathcal C[p_k,a] \circ C_i[a,p_j] \circ rev(P[p_k,p_j])$. Since $p_k$ is external to $C_i$, then $\mathcal C[p_k,a]$ is external to $C_i$. At vertex $a$, $\mathcal C$ enters $C'$. Observe that $\mathcal C[a,p_k]$ does not cross $P$ and needs to enter $P$ from the right before reaching $p_k$. This means that either $\mathcal C[a,p_k]$ exits $C'$ or it passes through $p_j$. However, 
        $\mathcal C[a,p_k]$ cannot exit $C'$
        at $\mathcal C[p_k,a]$ because $\mathcal C$ is a simple cycle, it cannot exit $C'$ at $P[p_k,p_j]$ because $\mathcal C$ crosses $P$ only once (at $p_k$), so if $\mathcal C[a,p_k]$ exits $C'$ it must do so at some vertex $b$ of $C_i[a,p_j]$. Since
        $C_i[a,p_j]$ does not cross $P$ at all, we can replace
        $\mathcal C[a,b]$ with $C_i[a,b]$ to get a globally minimum cycle that does not cross $C_i$ at $a$. The same argument shows that $\mathcal C[a,p_k]$ cannot pass through $p_j$.
       
       Before moving on,
        observe that we have just proven a stronger claim since we did
        not use the fact that $a$ is the {\em first} crossing
        vertex nor the fact that $p_k$ is {\em strictly} external to
        $C_i$ (i.e., the claim holds even if $p_k$ is on
        $C_i$). Namely, we proved that $(\ast)$ If $k <j$ and $k\leq \ell$ then
        $\mathcal C$ does not cross the separation finger.
	
	\item  If $k  \ge \ell$, then it must be that $p_k$ is enclosed by some $p_x$-to-$p_y$ finger $S'$ that is above $P$ (i.e., $S$ and $S'$ cross as in case I in Claim~\ref{claim:crossing}). Observe that $S'$ encloses only vertices of some single piece $R$.
	    Since $\mathcal C[p_k,\cdot]$ first intersects $S$ it must be that $S$ enters the finger $S'$ (at some vertex $c$) and exits the finger $S'$ (at some vertex $d$) such that $a\in S[c,d]$ (notice that it is possible that $a=c=d$). Furthermore, $\mathcal C[p_k,\cdot]$ first crosses $S[c,d]$ at $a$ and must eventually exit the finger $S'$ (at some vertex $b \in C_i[p_x,p_y]$).   
	  We first claim that $b\not \in C_i[p_x,d]$. This is because that would imply that for some $u\in S[a,d]$ the subpaths $\mathcal C[a,u]$ and $C_i[a,u]$ are two $a$-to-$u$ shortest paths that do not cross $P$. We can therefore replace $\mathcal C[a,u]$ with $C_i[a,u]$. 
	    Finally, we claim that $b\not \in C_i[d, p_y]$. 
	    This is because if $b \in C_i[d, p_y]$ then in $C_i$ we could replace $C_i[a,d] \circ C_i[d,d] \circ C_i[d,b]$ with $\mathcal C[a,a] \circ \mathcal C[a,b]$: (1) We are allowed to do such replacing because $\mathcal C[p_k,b]$ is enclosed by $S'$ and so is completely contained in $R$, and (2)   
	    This can only make $C_i$ shorter because $ C_i[d,d]$ is not longer than $\mathcal C[a,a]$ (since $\mathcal C[a,a]$ is the globally minimum cycle) and because $C_i[a,d] \circ C_i[d,b]$ is not longer than $\mathcal C[a,b]$ (since they are both contained in $R$ and do not cross $P$, and since $\mathcal C[a,b]$ is a globally shortest $a$-to-$b$ path). 
	    
	 \end{enumerate}
	 
\item{\em $S$ is a $p_x$-to-$p_y$ finger of $C_i$ below $P$.}	Since $p_k$ is external to $C_i$ and since $\mathcal C[p_k,\cdot]$ crosses $C_i$ first in a finger below $P$ we can conclude that $k < \ell$. 
Let $C'$ be the cycle $\mathcal C[p_k,a] \circ C_i[a,p_y] \circ rev(P[p_k,p_y])$. 
At vertex $a$, $\mathcal C$ enters $C'$ and must exit $C'$ before reaching $p_k$. $\mathcal C$ cannot exit at $\mathcal C[p_k,a]$ because $\mathcal C$ is simple, and it cannot exit at 
 $P[p_k,p_y]$ because $\mathcal C$ crosses $P$ only once, so $\mathcal C$ must exit $C'$ at some vertex $b$ of $C_i[a,p_y]$. However, since $C_i[a,p_y]$ does not cross $P$ at all, we can replace $\mathcal C[a,b]$ with $C_i[a,b]$. 

\begin{figure}[h!]
	\centering
	\includegraphics[scale=.5]{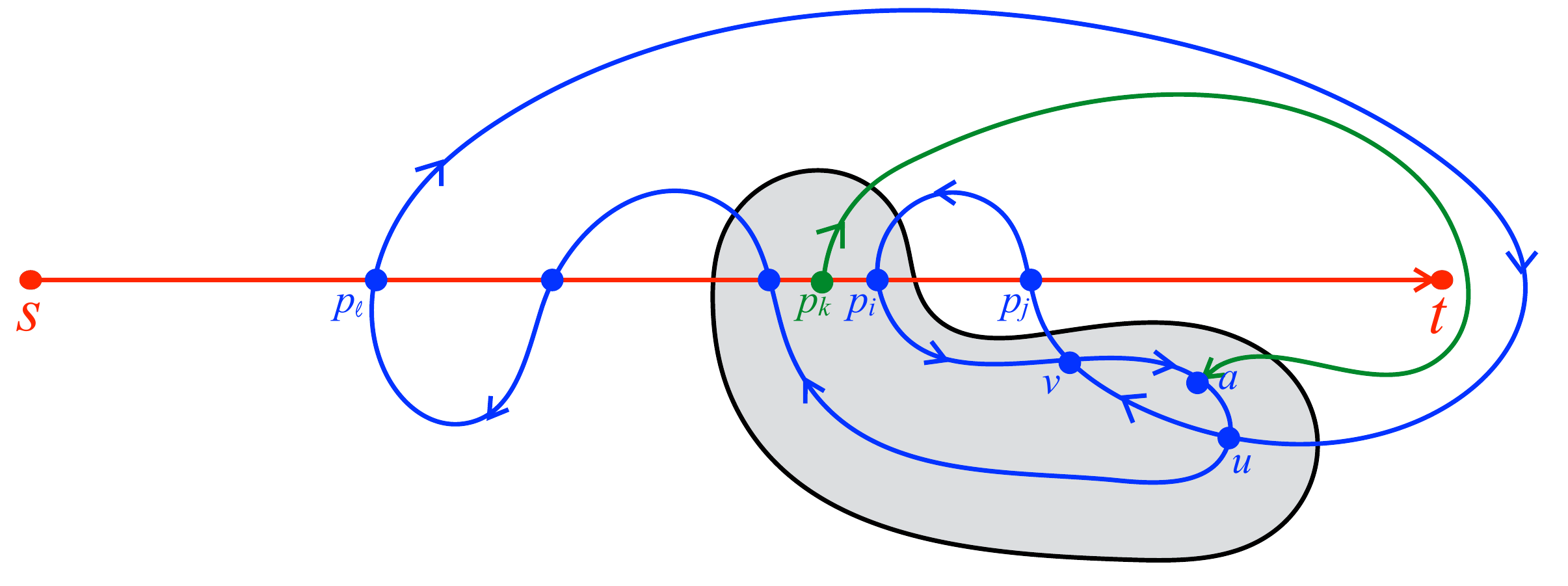}
	\caption{A finger of $C[p_i,p_\ell]$ below $P$ that crosses
          the separation finger.}
	\label{fig:Correctness3}
\end{figure}

\item {\em $S$ is a $p_x$-to-$p_y$ finger of $C_i$ above $P$.} 
	
	\begin{enumerate}
	  \item	\label{case3.1} If $k < \ell$, then after crossing at $a$ the
          cycle $\mathcal C$ enters the finger $S$ (otherwise
          $\mathcal C$ crosses $P$ more than once). In order for this
          to happen, the finger $S$ must cross the separation finger
          (as in case I in Claim~\ref{claim:crossing} and as illustrated in  Figure~\ref{fig:Correctness}). 
To see why, consider the cycle  $C'= C_i[p_\ell, p_j] \circ
rev(P[p_\ell,p_j])$ ($C'= C_i[p_\ell, p_j] \circ
P[p_j,p_\ell]$) if $\ell \le j$ (if $j<\ell$). Since $p_k$ is strictly
external to $C'$, $\mathcal C$ cannot enter $C'$ before crossing the finger $S$. Hence the finger $S$ must cross the separation finger. 
 This means that $i < y < x \leq j$ because if  $\ell < y < x <i$ and
 $S$ crosses the separation finger at node $u$ then we can throw
 $C_i[u,u]$ from $C_i$ and obtain a shorter cycle $C_i$ that passes
 through $p_i$ and separates $s$ and $t$. Since $i < y < x \le j$
 then, by Observation~\ref{observation:above}, 
 the finger $S$ is entirely contained in a single piece $R$ and
 encloses no holes. 
Recall that $b_1$ and $b_2$ are the boundary vertices preceding and
following $a$ on $\mathcal C$, and by definition belong to different
sides of $C_i$. Thus $b_1,b_2$ are boundary vertices of $R$.
We claim that the DDG edge $b_1b_2$ belong to the same side of $C_i$,
which is a contradiction.
 This is because 
for $b_1$ and $b_2$ to belong to different sides of $C_i$,  $\mathcal C[b_1,b_2]$ must either cross $P$ from left to right (but $\mathcal C$ cannot do this by definition) or cross the separation finger (but $\mathcal C$ cannot do this by $(\ast)$). 
  
	\item	If $k\geq \ell$ then $p_k$ is enclosed by some finger $S'$ of $C_i$ above $P$. Note that $S'$ can be either the $p_x$-to-$p_y$ finger $S$ (and then $y<k<x$), or some other $p_w$-to-$p_z$ finger (i.e., $S$ and $S'$ cross as in case III of Claim~\ref{claim:crossing} as illustrated in Figure~\ref{fig:Correctness2}). Since $p_k$ is enclosed by the $S'$ finger, $\mathcal C[p_k,\cdot]$ must exit $S'$ at some vertex $a'$ (note that if $S'=S$ then $a'=a$). Then, before reaching $a'$
          again, $\mathcal C[a',\cdot]$ must cross the separation finger (at some
          vertex $b$). 
          
         If $C_i[a',b]$ does not include $p_i$, then since $\mathcal C[a',b]$ does not cross $P$ we should replace in $C_i$ the subpath
          $C_i[a',b]$ with $\mathcal C[a',b]$. If $C_i[a',b]$ does include $p_i$, then $\mathcal C[p_k,a']$ is entirely contained in a single piece $R$ and so  
          in $C_i$ we should replace the subpath
          $C_i[b,a']$ with $\mathcal C[b,a']$. 

	\begin{figure}[h!]
	\centering
	\includegraphics[scale=.6]{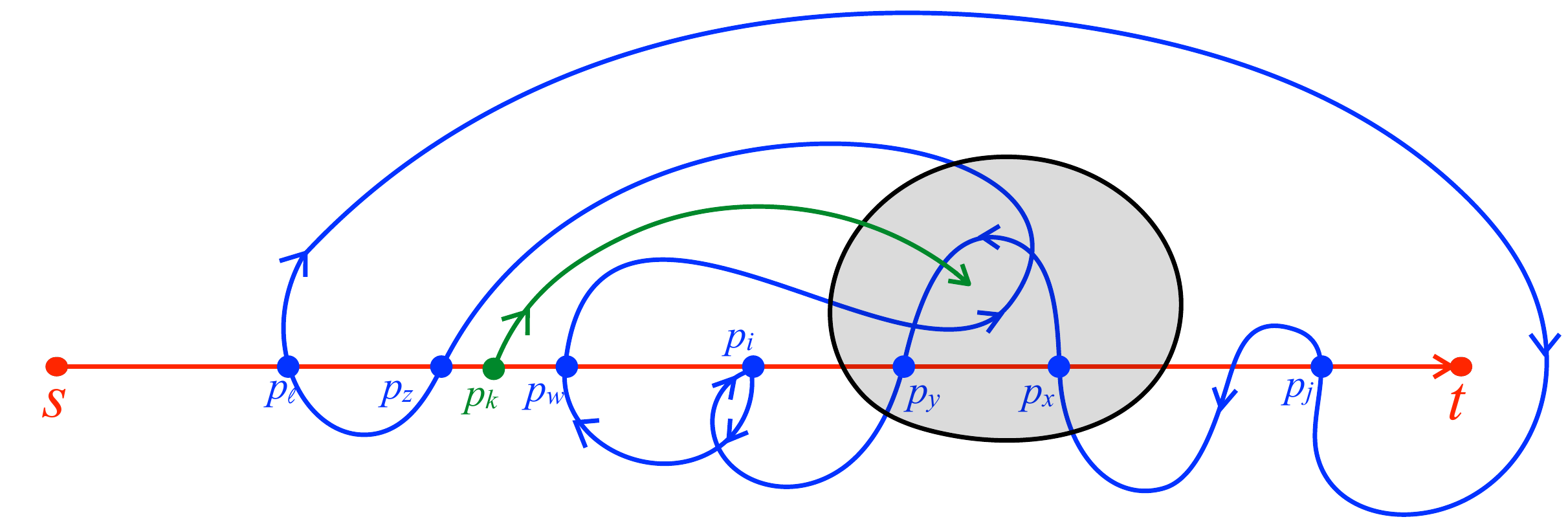}
	\caption{Crossing $p_x$-to-$p_y$ and $p_w$-to-$p_z$ fingers of
          $C_i$ (in blue) above $P$. Notice that $\mathcal C$ (in
          green, shown partially) first crosses the $p_x$-to-$p_y$ finger even though $z<k<w$. We prove that such crossings cannot occur.}
	\label{fig:Correctness2}
\end{figure}

	\end{enumerate}

\end{enumerate}	

\noindent Next we consider the case where $p_k$ is internal to $C_i$ or lies on $C_i$. Again, the finger $S$ is one of three types:
\begin{enumerate}[resume,leftmargin=*]

\item \label{case4} {\em $S$ is the separation finger.} In this case, if $k\leq\ell$ then since $p_k$ is internal to $C_i$ it means that $j<\ell$, 
$C_i[p_j,p_i]= P[p_j,p_i]$, and $p_k$ lies on $C_i$. 
Furthermore, $C_i[p_\ell,p_k]$ (and hence also $C_i[a,p_k]$) does not cross $P$ at all. This means that in  $\mathcal C $ we can replace $\mathcal C[a,p_k]$ with $C_i[a,p_k]$ to get a globally minimum cycle that does not cross $C_i$ at $a$. 
If on the other hand $k\ge \ell$, then $\mathcal C$ exits $C_i$ in $a$ and $\mathcal C(a,p_k]$ must intersect $C_i$ (because $p_k$ is either in or on $C_i$). Let $b$ be the last vertex where  $\mathcal C(a,p_k]$ intersects $C_i$. 
	If $b\in C_i[a,p_i)$ then in $C_i$ we could replace $C_i[a,b]$ with $\mathcal C[a,b]$ since $\mathcal C[a,b]$ does not cross $P$.
	If on the other hand $b\in C_i[p_i,a)$ then $C_i[a,b]$ (as opposed to $\mathcal C[a,b]$) is required to pass through $p_i$. In this case we therefore consider $C_i[b,a]$ and $\mathcal C[b,a]$: even though $\mathcal C[b,a]$ crosses $P$ (at $p_k$) the subpath $\mathcal C[b,p_k]$ is entirely contained in a single piece $R$ (because it is enclosed by a  finger of $C_i[p_i,p_\ell]$ below $P$, which by Observation~\ref{observation:below} is contained in a single piece). Therefore, in $C_i$ we could replace 
	$C_i[b,a]$  with $\mathcal C[b,a]$. 
		
\item {\em $S$ is a $p_x$-to-$p_y$ finger of $C_i$ below $P$.} 

\begin{enumerate}
	\item If $k=\ell$ then if $a$ is on $C_i[p_j,p_i]$ then we can replace $\mathcal C[p_k,a]$ with $C_i[p_k,a]$. If  $a$ is on $C_i[p_i,p_\ell]$ then we can replace $\mathcal C[a,p_k]$ with $C_i[a,p_k]$. 
	We conclude that $k\ne \ell$. 
	
	\item  If $j<i$ then $C_i[p_j,p_i]= P[p_j,p_i]$ so Observation~\ref{observation:below} applies to $S$.  
	\begin{enumerate}
	
	\item If $k \le y$ or $k\ge x$,  then after entering the
          finger $S$ at $a$ and before reaching $p_k$ again, $\mathcal
          C[p_k,\cdot]$  must exit $S$ at some vertex $b$. However then, by Observation~\ref{observation:below},  
          there is a DDG edge whose corresponding path contains $\mathcal C[a,b]$ as a subpath and has both endpoints on the same side of $C_i$.

\item \label{case5.3} If $y<k<x$ and $j\le k$, then $k$ is at the base of the finger $S$. Since $j<i$ we have that the bases of all fingers are in $P[p_\ell,p_i]$. This implies that $k \le i$. Let $b$ be the last
  vertex of $\mathcal C$ strictly before $p_k$ that belongs to $S$ (see Figure~\ref{fig:5.3} which falls under case II of Claim~\ref{claim:crossing}). (i) $p_k$ is a vertex of $C_i$, (ii) $\mathcal C[b,p_k]$ is enclosed by $S$, (iii)  $\mathcal C[b,p_k]$ is contained in a single piece $R$, (iv) $\mathcal C[p_k,b]$ does not cross $P$. 
This implies that $C_i$ is a non-simple cycle in which we could replace $C_i[b,p_k]$ with the non-simple cycle $\mathcal C \circ \mathcal C[b,p_k]$.   
	    To see that this can only make $C_i$ shorter define $u$ to 
            be the last vertex of $C_i[p_\ell,p_j)$ that belongs to
            $S$ if $j>y$ and $p_y$ otherwise. 
            Observe that $C_i[b,p_k]$ can be decomposed as $C_i[b,u] \circ C_i[u,p_\ell] \circ C_i[p_\ell,u] \circ C_i[u,p_k]$.
            The cycle $C_i[u,p_\ell] \circ C_i[p_\ell,u]$ is not longer than $\mathcal C$ (since $\mathcal C$ is globally minimum), and $C_i[b,u]\circ C_i[u,p_k]$ is not longer than $\mathcal C[b,p_k]$ (since they are both contained in $R$, and since $\mathcal C[b,p_k]$ is a globally shortest $b$-to-$p_k$ path). 

	\begin{figure}[h!]
	\centering
	\includegraphics[scale=.6]{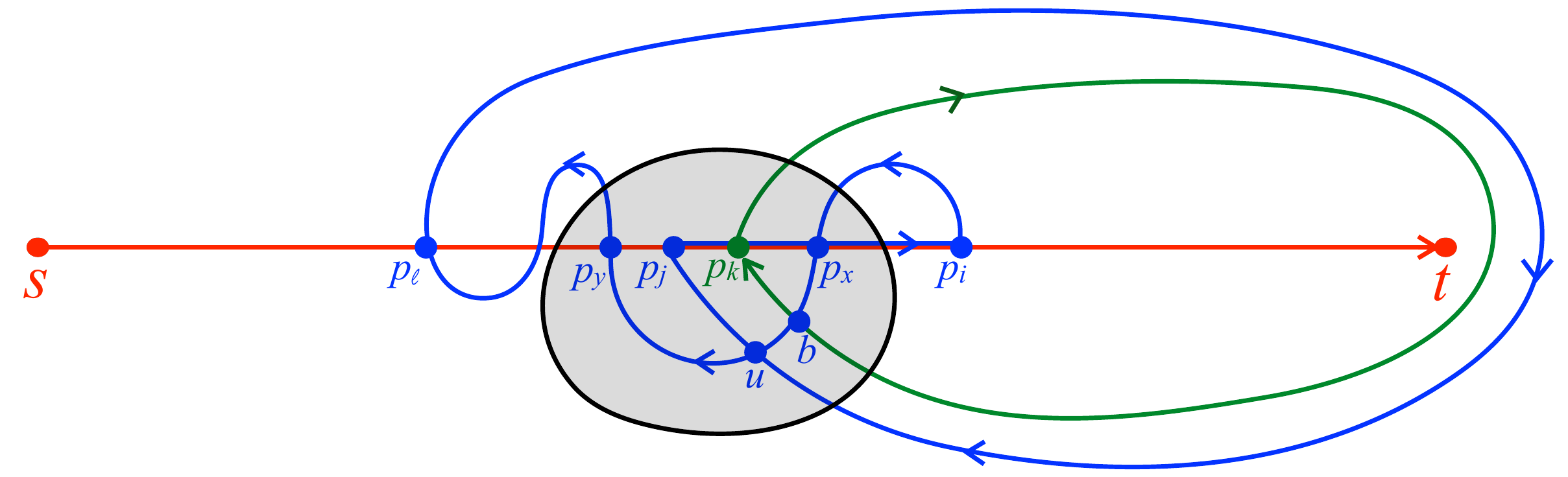}
	\caption{$\mathcal C$ (in green) crosses the $p_x$-to-$p_y$ finger of $C_i$ (in blue) at $b$. Such crossings cannot occur because the (non-simple) subpath $C_i[b,p_k]$ can be made shorter by replacing it with $\mathcal C\circ \mathcal C [b,p_k]$.}
	\label{fig:5.3}
\end{figure}

\item If $y<k<x$ and $k<
j$, then again, define $u$ to 
            be the last vertex of $C_i[p_\ell,p_j)$ that belongs to
            $S$. $\mathcal C$ enters at $a$ the simple cycle 
$\mathcal C[p_k,a] \circ C_i[a,u] \circ C_i[u,p_j] \circ
rev(P[p_k,p_j])$ (where $C_i[u,p_j]$ is the simple path in $C_i$
between $u$ and $p_j$).
It must exit this cycle in order to get to $p_k$ from below. 
It cannot cross at $P$ by definition of $p_k$.
It cannot leave the cycle through $\mathcal C$ because of simplicity.
It cannot leave the cycle through $C_i[a,u]$ because $C_i[a,u]$ and
$\mathcal C[a,u]$ are two different shortest paths.  If it leaves the
cycle at a vertex $v$ of $C_i[u,p_j]$ then, similarly to
case~\ref{case5.3}, we can make $C_i$ shorter by replacing $C_i[a,v]$
with $\mathcal C \circ \mathcal C[a,v]$. 
\end{enumerate}
\item If $j\ge i$ then $k>\ell$. This means that $S$ is a finger of $C_i[p_i,p_\ell]$ below $P$ that crosses the separation finger (see Figure~\ref{fig:Correctness3} which falls under case II of Claim~\ref{claim:crossing}). In other words, $a$ is enclosed by the cycle $C_i[p_\ell,p_j]\circ rev(P[p_\ell,p_j])$ since otherwise $\mathcal C[p_k,\cdot]$ must first exit this cycle which would mean that $S$ is the separation finger (and not a finger below). 
\begin{enumerate}
\item \label{5.3.1} If $k < y$ then $\mathcal C[p_k,\cdot]$ enters the cycle 
$\mathcal C[p_k,a] \circ C_i[a,p_y]$ at vertex $a$ and must exit this cycle before reaching $p_k$ again. It cannot exit at $\mathcal C[p_k,a]$ because $\mathcal C$ is simple and it cannot exit $C_i[a,p_y]$ at any vertex $b$ because that would imply that $C_i[a,b]$ and $\mathcal C[a,b]$ are two $a$-to-$b$ shortest paths. 
  
\item If $x<  k < j$ then $p_k$ is on the base of some bottom finger $S' \ne S$. To reach $p_k$ again, $\mathcal C[p_k,\cdot]$ must cross the separation finger and then cross $S'$. We have already proved in case~\ref{case4} that this is impossible.

\item If $y\le k \le x$ then let $u$ and $v$ be the vertices that belong to both $S$ and the separation finger and are the endpoints of their intersecting subpaths (see Figure~\ref{fig:Correctness2}). At vertex $a$, $\mathcal C[p_k,\cdot]$ enters the cycle $\mathcal C[p_k,a] \circ C_i[a,p_y]$ and we have already seen in case~\ref{5.3.1} that $\mathcal C[a,p_k]$ must remain inside this cycle. This implies that $\mathcal C[a,\cdot]$ must cross $C_i[u,p_j]$ before reaching $p_k$. Let $b$ denote the last such crossing vertex. In $C_i$, we can therefore replace $C_i[a,u]\circ C_i[u,u]\circ C_i[u,b]$ with $\mathcal C[a,a]\circ \mathcal C[a,b]$ to obtain a shorter cycle. This is because $\mathcal C$ is the globally shortest cycle and so $\mathcal C[a,a]$ is shorter than $C_i[u,u]$ and $\mathcal C[a,b]$ is shorter than $C_i[a,u]\circ  C_i[u,b]$.
 
\item If $k\ge j$ then $\mathcal C[p_k,\cdot]$ cannot cross the separation finger: If it crosses at vertex $b$ then it exists the cycle $\mathcal C[p_k,b] \circ C_i[b,p_j]$ and it must enter this cycle again before reaching $p_k$. However, it cannot enter at $\mathcal C[p_k,b]$ because $\mathcal C$ is simple and it cannot enter at a vertex $c$ of $C_i[b,p_j]$ because then $\mathcal C[b,c]$ and $C_i[b,c]$ are two $b$-to-$c$ shortest paths. This means that $\mathcal C(a,\cdot]$ can only cross $S$ (an odd number of times). However, since $S$ is a simple finger it is contained in a single piece $R$ (by Observation~\ref{observation:below}). Such crossings are available in the DDG even after cutting along $C_i$.  

\end{enumerate}
   
\end{enumerate}
	
\item {\em $S$ is a $p_x$-to-$p_y$ finger of $C_i$ above $P$.} 

\begin{enumerate}
        \item If $k=\ell <j$ then, if the first edge of $\mathcal C$
          that leaves $C_i$  after $p_k$ is enclosed by the separation finger,
          then $\mathcal C$ first crosses the separation finger, so we
          are in case~\ref{case1.2}. If the first edge of $\mathcal C$
          that leaves $C_i$  after $p_k$ is not enclosed by the separation finger,
          then $S$ must also cross the separation finger, so we are in
          case~\ref{case3.1}.
	\item If $k< j$, then $\mathcal C[p_k,\cdot]$ must exit
          the cycle $C_i[p_\ell,p_j]\circ rev(P[p_\ell,p_j])$ (at some
          vertex $a'$ on the separation finger $C_i[p_\ell,p_j]$) and
          then (since $p_k$ is enclosed by $C_i$) intersect $C_i$
          again (in this case $p_k$ is on the base of a bottom finger,
          so the last vertex $b$ of $C_i$ intersected by $\mathcal C$
          before reaching $p_k$ belongs to this bottom finger). 
As in case~\ref{case4} above, 
if $b\in C_i[a',p_i)$  we could replace $C_i[a',b]$ with $\mathcal C[a',b]$ and if  $b\in C_i[p_i,a')$  we could replace $C_i[b,a']$  with $\mathcal C[b,a']$.  

\item \label{case6.2} If $i\le j\le k$, then the finger $S$ must be such that $x\le k$. 
If $S$ is a finger of $C_i[p_j,p_i]$ (i.e., $i < y < x \leq j$), then
$\mathcal C[p_k,\cdot]$ enters the finger at $a$ and must exit the
finger (at some vertex $b\in C_i[p_x,p_y]$) before reaching
$p_k$. 
Observation~\ref{observation:above}, 
 the finger $S$ is entirely contained in a single piece $R$ and
 encloses no holes. 
Therefore, the boundary vertices $b_1$ and $b_2$ preceding and
following $a$ on $\mathcal C$ belong to the same side of $C_i$,
contradicting their definition.
 If on the other hand $S$ is a finger of $C_i[p_i,p_\ell]$ (i.e., $\ell < y < x <i$), 
then first edge of $\mathcal C[a,\cdot]$ is enclosed by the cycle $\mathcal C[p_k,a]\circ C_i[a,p_y]\circ P[p_y,p_k]$. Thus $\mathcal C[a,p_k]$ must exit this cycle before reaching $p_k$. It can only exit at some vertex $b\in C_i[a,p_y]$. However, this implies that $C_i[a,b]$ and $\mathcal C[a,b]$ are two shortest $a$-to-$b$ paths that do not cross $P$.   

\item If  $j\le i \le k$,
then as in case~\ref{case6.2}, this means that $\mathcal C[p_k,\cdot]$ at vertex $a$ enters the cycle $\mathcal C[p_k,a]\circ C_i[a,p_y]\circ P[p_y,p_k]$ and must exit this cycle at some vertex $b\in C_i[a,p_y]$. Implying that $C_i[a,b]$ and $\mathcal C[a,b]$ are two shortest $a$-to-$b$ paths that do not cross $P$.  

\item If  $j \le k\le i$,  then $p_k$ lies on $C_i[p_j,p_i]=
  P[p_j,p_i]$, and $S$ is a finger of $C_i[p_i,p_\ell]$.  This means
  that in $C_i$ we could replace $C_i[a,p_k]$ with $\mathcal C[a,p_k]$
  because $C_i[a,p_k]$ does not visit $i$, and $\mathcal C[a,p_k]$ does not cross $P$.
  
\end{enumerate}

\end{enumerate}	

Since $p_k$ might not be a boundary vertex of the $r$-division, we
must also argue that $p_k$ is assigned to the subgraph of the DDG in
which we have shown $\mathcal C$ exists (i.e., either $\DDG_s$ or $\DDG_t$).
Let $p_{first}$ and $p_{last}$ denote the first and last
vertices of $P$ that are also vertices of $C_i$. 
We have already seen that, in the cases where $\mathcal C$ crosses
$C_i$, $p_k$ either appears before $p_{first}$ or after $p_{last}$ on $P$. 
If, on the other hand
$\mathcal C$ does not cross $C_i$ at all then, since $\mathcal C$ crosses $P$ once,
$p_k$ cannot be inside a finger. Therefore, either $p_k$ appears on $P$ before
$p_{first}$ (and hence also before $p_i$), or  $p_k$ appears on $P$ after
$p_{last}$ (and hence also after $p_i$).

\subsection{Additional details about the use of the $\pmb{\mathbb
  Z_2}$-homology cover}\label{sec:more-cover}

When information about a DDG edge $uv$ of $Q$ is required during the
execution of the algorithm,
we want
it to correspond to a shortest $u$-to-$v$ path in $R$ that satisfies
Property~\ref{prop:simple-finger}. The appropriate shortest path is
represented in the MSSP data structure for the $\mathbb Z_2$-homology
cover of $R$ with $H$ being the subset of holes of $R$ that are not
holes of $Q$. This subset $H$ can be associated with $Q$ at the time
the implicit incision along $P$ is made. We need to be able to
infer the appropriate label of the vertex $v$. 
Among the (DDG) edges of $P$ that form the boundary of $Q$ there is a
constant number of edges that split the holes in $H$ (this is because
each such edge defines a distinct subpiece of $R \setminus Q$ that
contains at least one hole). When $R$ is implicitly cut along $P$, we mark the endpoints of such edges and store them
using a data structure that supports fast predecessor search ((i) given a
boundary vertex of $P$ find its marked predecessor on $P$, and (ii) given a
boundary vertex of $Q$ find its marked predecessor on the external
boundary of $Q$). For each pair of marked vertices $x,y$ we store the
crossing parity of an $x$-to-$y$ path in $Q$ with every
$A_\ell$. These crossing parities can be computed using the
information stored in the MSSP data structure for the (DDG) edges of
$P$. Whenever the appropriate label for $v$ for a DDG edge $uv$ is
required, it is obtained by querying the label stored for 
the pair of predecessors of $u$ and $v$. 

\section{The Division-Edge technique}\label{sec:division-edges}
A technique of \L\k{a}cki and Sankowski that we use
without change in our algorithm is the use of a {\em recursion graph} and {\em division
  edges} to efficiently keep track of the partition of the 
graph into subgraphs along the execution of the algorithm.
Since most
of the algorithm is run on the DDG rather than on the underlying
planar graph $G$, it is necessary to be able to quickly determine how to
partition the boundary vertices when separating a graph into the
subgraphs enclosed and not enclosed by some cycle of edges in the
DDG. Note that this is particularly challenging since, in general,
boundary vertices of a piece belong to multiple holes. 
The recursion graph stores the information required to perform this task.

For each piece $R$ of the $r$-division with external hole $h$ and
a constant number of internal holes $\{h_i\}$, we fix an arbitrary set of mutually
noncrossing $h$-to-$h_i$ paths $K_{h_i}$. We store for every edge $e$ of the DDG
of $R$ the crossing parity number of the path corresponding to $e$ and
each $K_{h_i}$. This information can be computed and stored
within the same bounds required to compute and store the DDG.

The recursion graph is a planar
embedded graph whose vertices are the boundary vertices of the
$r$-division of the input planar graph. Initially, the only edges
in the recursion graph are edges between consecutive boundary vertices
of the $r$-division that lie on the same hole of their piece. The
embedding of this initial graph is inherited from the embedding of the
input planar graph. Edges between consecutive vertices that do  not exist in the original graph are embedded along the corresponding subpath of the hole. 
Edges are added to the recursion graph when the
algorithm separates the graph into internal and external parts with
respect to some cycle $C$ of edges in the DDG. 
For each DDG edge $e$
of $C$, we add an edge $e'$ to
the recursion graph. The edge $e'$ has the same endpoints as $e$, and is embedded in the recursion graph
so that the crossing parity of each $K_{h_i}$ and the curve on the plain
that corresponds to $e'$ matches the parity stored for the DDG edge $e$. 
This guarantees that the partition of boundary vertices into the internal and external
subgraphs with respect to the DDG cycle $C$ is the same as the 
partition of the vertices with respect to the corresponding cycle in
the recursion graph. Since the number of vertices and edges of the
recursion graph is linear in the number of boundary vertices,
computing the partition of a subgraph $G'$ takes linear time in the number of boundary
vertices in $G'$.

\section{Graphs Embedded on a Surface} \label{sec:genus}

In this section we briefly describe a generalization of our algorithm for finding the shortest cycle in a graph embedded on a surface with a bounded genus $g$.
 We present two algorithms for the problem, one runs in $O(g^2 n \log n)$ time with high probability, and the other runs in $O(g n \log^2 n)$ in the worst case. Notice that the duality between cuts and cycles does not hold for
$g>0$, so these algorithms do not find the minimum cut in such graphs.

The first algorithm uses a \emph{greedy system of loops}~\cite{EW05} based at an arbitrary vertex $o$. This is a set $L$ of $2g$ undirected cycles in $G$, each of them containing the basepoint $o$, such that every undirected cycle $S$ in $L$ consists of an edge $uv$, a shortest $o$-to-$v$ path and, a shortest $o$-to-$u$ path. This is similar to a shortest path separator. If we make incisions in $G$ along the paths that define the cycles of $L$, we remain with a planar graph $G_L$.

We begin by finding a greedy system of loops $L$ in $O(n)$ time using the algorithm of Erickson and Whittlesey~\cite{EW05}. The shortest cycle in $G$ is either a cycle in $G_L$, or crosses one of the undirected cycles of $L$. We find the shortest cycle in $G_L$ in $O(n \log \log n)$ time using the algorithm for planar graphs (to get this time bound we assume that $g = o(\sqrt{n})$, since otherwise the second algorithm which we present next is faster). We find the shortest cycle if it crosses a member of $L$ using an MSSP algorithm, similarly to the implementation of Reif's algorithm~\cite{R83} which we described in Section~\ref{sec:conquerstep}.
We use here the fact that a cycle of $L$ is composed of two shortest paths and that Lemma~\ref{lem:commonsubpath} does not depend on the planarity of the input graph. We apply $O(g)$ times the MSSP algorithm of Cabello et al.~\cite{CCE13}. This takes $O(g^2 n \log n)$ time with high probability. The shortest cycle in the graph is the shortest among the $O(g)$ cycles that we find for each member of $L$ and the shortest cycle in $G_L$.

Our second algorithm for embedded graphs uses a \emph{planarizing set}, which is a set of edges or vertices whose removal from the graphs leaves a planar graph. We begin by finding a planarizing set $R$ of $O(\sqrt{gn})$ edges in time linear in the number of the edges of the graph~\cite{DV95}. We remove the edges of $R$ and get a planar graph $G_R$. We find the shortest cycle in $G_R$ using our algorithm for planar graphs. For each edge $uv$ of $R$ we find the shortest cycle containing $uv$ by computing the shortest $v$-to-$u$ path. This takes $O(\sqrt{gn} \log^2 n)$ time per edge of $R$ after $O(n \log^2 n)$ time preprocessing using a variant of FR-Dijkstra~\cite{FR06}, as noted by Smith. This gives a total running time of $O(g n \log^2 n)$ (we assume here that $g = o(n)$, otherwise it is simple to get this time bound). The shortest cycle in $G$ is the shortest among the shortest cycle containing edges of $R$ and the shortest cycle in $G_R$.

\newpage
\bibliographystyle{plain}

\end{document}